\documentclass[]{pasj01}
\usepackage{lineno}
\Received{$\langle$2022 Jul. 15$\rangle$}
\Accepted{$\langle$2022 Aug. 22$\rangle$}
\Published{$\langle$publication date$\rangle$}
\SetRunningHead{K.Murata \& T.T.Takeuchi}{Deblurring galaxy images}
\usepackage{multirow}
\usepackage{url}
\usepackage{threeparttable}

\begin{document}

\title{ Deblurring galaxy images with Tikhonov regularization on magnitude domain}
\author{
  Kazumi \textsc{Murata},\altaffilmark{1,}\footnotemark[*] and
  Tsutomu T. \textsc{Takeuchi}\altaffilmark{2,3}
}

\altaffiltext{1}{
  National Astronomical Observatory of Japan,  2-21-1 Osawa, Mitaka, Tokyo 181-8588, Japan
}
\altaffiltext{2}{
  Division of Particle and Astrophysical Science, Nagoya University, Furo-cho, Chikusa-ku, Nagoya 464-8602, Japan
}
\altaffiltext{3}{
  The Research Center for Statistical Machine Learning, the Institute of Statistical Mathematics, 10-3 Midori-cho, Tachikawa, Tokyo 190-8562, Japan
}

\email{kazumi.murata@nao.ac.jp}


\maketitle

\begin{abstract}
  We propose a regularization-based deblurring method that works efficiently for galaxy images. 
  The spatial resolution of a ground-based telescope is generally limited by seeing conditions and much worse than space-based telescopes. 
  This circumstance has generated considerable research interest in restoration of spatial resolution. 
  Since image deblurring is a typical inverse problem and often ill-posed, solutions tend to be unstable.
  To obtain a stable solution, much research has adopted regularization-based methods for image deblurring, but the regularization term is not necessarily appropriate for galaxy images. 
  Although galaxies have an exponential or S{\'e}rsic profile, the conventional regularization assumes the image profiles to behave linear in space. 
  The significant deviation between the assumption and real situation leads to blurring the images and smoothing out the detailed structures. 
  Clearly, regularization on logarithmic, i.e. magnitude domain, should provide a more appropriate assumption, which we explore in this study. 
  We formulate a problem of deblurring galaxy images by an objective function with a Tikhonov regularization term on magnitude domain. 
  We introduce an iterative algorithm minimizing the objective function with a primal-dual splitting method. 
  We investigate the feasibility of the proposed method using simulation and observation images. 
  In the simulation, we blur galaxy images with a realistic point spread function and add both Gaussian and Poisson noises. 
  For the evaluation with the observed images, we use galaxy images taken by the {\it Subaru} HSC-SSP. 
  Both of these evaluations show that our method successfully recovers the spatial resolution of the images and significantly outperforms the conventional methods.
  The code is publicly available from the GitHub(\url{https://github.com/kzmurata-astro/PSFdeconv_amag}).
\end{abstract}

\section{Introduction}
Images with high spatial resolution have brought significant impacts into astronomical research. 
Sharp images provided by the Hubble Space Telescope {\it (HST)} enable detailed analyses such as the morphological classification of galaxies, size measurement, and spatially resolved analysis (e.g. Shibuya et al. 2015, Suess et al. 2019ab). 
Further, higher resolution images provided by forthcoming space missions would also revolutionize studies of galaxy evolution and formation. 
\par

In contrast to these space telescopes, the spacial resolution of ground-based telescopes is severely limited by the sky-seeing. The atmospheric turbulence leads to a large point spread function (PSF) and degrades the spatial resolution. 
Nonetheless, the large ground-based telescopes such as {\it Subaru} still have many advantages such as obtaining a large amount of photons and providing wide field surveys (e.g. {\it Subaru} HSC-SSP; Aihara \yearcite{2018PASJ...70S...4A}, \yearcite{2019PASJ...71..114A}). 
Clearly, improvements of the spatial resolution of the ground-based telescopes would also provide significant impacts in astronomy.
\par

In this study, we focus on a software based method, i.e. PSF deconvolution.
While a hardware based technique such as adoptive optics (AO) has provided significantly sharp images (e.g. Suzuki et al. \yearcite{2019PASJ...71...69S}), the seeing effects could not be completely corrected in general.
Furthermore, even AO imaging is limited by the diffraction limit and has a PSF. 
As a consequence, the spatial resolution of even AO imaging can be improved by a PSF deconvolution technique.
\par

The PSF deconvolution is a technique that restores images degraded by a PSF and significantly improves the spatial resolution.
There has been much research investigating deconvolution methods 
\citep{2002PASP..114.1051S, Anconelli04, 2008ApJ...675.1304R, 2021ApJS..257...66C, 2022PASJ...74...73S}, 
including blind deconvolution \citep{Shi17, 2020MNRAS.496.4209F,2022ApJ...926...88H}
and machine-learning based methods \citep{2017MNRAS.467L.110S, 2020A&A...641A..67S, 2021arXiv210309711G, 2022arXiv220307412N}.
The most famous and classical method is the Richardson-Lucy algorithm (RL; Richardson \yearcite{1972JOSA...62...55R}, Lucy \yearcite{1974AJ.....79..745L}), which assumes the pixel values to follow a Poisson distribution and would provide a maximum likelihood solution. 
\par

The PSF deconvolution is an inverse problem and often ill-posed, leading to unstable solutions. 
Especially in case of restoration of noisy images, the noise would be significantly amplified. 
Hence, when using the RL algorithm, the iteration must be stopped at a suitable step, where the number of iterations can be interpreted as a hyper-parameter. 
In contrast, introducing a regularization term into the objective function allows a stable and more appropriate solution. 
Widely known regularization terms are, Tikhonov regularization (quadratic norm with a linear operator), total variation (TV; Rudin et al.\yearcite{1992PhyD...60..259R}), maximum entropy method (MEM; Narayan \& Nityananda \yearcite{1986ARA&A..24..127N}), and sparsity in wavelet transformation \citep{2007ITIP...16..297S}.
These regularization-based methods significantly contribute to the astronomical field, and recent research from the event horizon telescope (EHT) project reports reconstruction of black hole shadow images \citep[Akiyama et al. 2022]{2014PASJ...66...95H,2017AJ....153..159A,2019ApJ...875L...1E}. 
\par

Nonetheless, these regularization terms are not necessarily appropriate for galaxy images. 
Although regularization such as the TV and Tikhonov adopts a linear operator when calculating the norm, galaxy profiles are much steeper than linear and generally expressed by an exponential function or S{\'e}rsic profile \citep{1963BAAA....6...41S}.
In other words, the image gradient is so steep that the conventional regularization smooths out the high-contrast structures. 
As a natural consequence, regularization on logarithm (i.e. magnitude) image domain should alleviate the problem. 
\par

In this work, we propose a PSF deconvolution method with an appropriate regularization term for deblurring galaxy images. 
We introduce an objective function with a Tikhonov regularization term on magnitude domain, and optimize it via a primal-dual splitting method \citep{Condat2013}. 
Evaluations with simulation and observation images show that our method outperforms the conventional methods. 
The rest of this paper is organized as follows.
In section 2, we provide the objective function to be solved, the optimization method, and comparison methods.
In section 3, we evaluate our methods with both simulation and observation images. 
In section 4, we show the effectiveness of our methods and discuss some limitation and future prospects. 
Finally, we conclude our contribution in section 5.

\section{Methods}
\label{method}
In general, PSF deconvolution is an ill-posed problem due to observational noises, which lead to unstable solutions. 
The most basic way to restore an original image $\mathbf{X}$ from an observational image $\mathbf{Y}$ is a least-$\chi^2$ method:
\begin{equation}
  \label{eq:lschi2}
  \mathbf{X} = \mathop{\arg\min}_\mathbf{X} \left[ \frac{1}{2} \chi^2(\mathbf{X}) \right]\; s.t\; \mathbf{X} > 0,
\end{equation}
\begin{equation}
  \label{eq:chi2}
  \chi^{2}\left({\bf X}\right) = \left({\bf AX} - {\bf Y}\right)^{\top} {\bf \Sigma^{-1}} \left({\bf AX} - {\bf Y}\right),
\end{equation}
where $\boldsymbol{\Sigma}$ is a diagonal matrix whose elements are variances $\sigma^2_{y}$ at the corresponding position and $\mathbf{A}$ is a system matrix to convolve the PSF with the original image.
The positivity constraint is added because pixel values should not be negative. 
\par

However, this method is highly sensitive to observational noise.
To address this problem, many previous studies adopt a regularization term to the objective function \citep{Idier};
\begin{equation}
  \label{eq:cost0}
        \mathop{\arg\min}_\mathbf{X} \left[
          \frac{1}{2} \chi^{2}\left(\mathbf{X}\right) + \lambda R\left(\mathbf{X}\right)
          \right] \; s.t.\; \mathbf{X} > 0, 
\end{equation}
where $R$ is a regularization function and $\lambda$ is the regularization parameter to balance the data fidelity term ($\chi^2$ term in this case) and the regularization term.
The regularization term works as {\it a priori} probability distribution function in the framework of Bayesian inference, and should be appropriately chosen for images to be restored. 
\par
The quality of the restored images strongly depends on the regularization term in equation \ref{eq:cost0}.
In the following, we propose a regularization term appropriate for galaxy images.

\subsection{Tikhonov regularization on magnitude domain}
Natural images such as astronomical images are known to have smooth structures.
In other words, values of adjacent pixels are similar to each other and norms of the image gradient should be small. 
This knowledge can be {\it a priori} information and often adopted as regularization terms, e.g. TV and Tikhonov regularization.
The Tikhonov regularization is a quadratic norm with a linear operator; 
\begin{eqnarray}
  \label{eq:tikreg}
  R\left({\bf X}\right)
  &=& \|{\bf LX}\|^{2} \\
  &=& \sum_{i,j}\left\{ \left( x_{i,j} - x_{i+1,j}\right)^2 + \left( x_{i,j} - x_{i,j+1}\right)^2\right\},
\end{eqnarray}
where we adopt {\bf L} as a differential operator for both horizontal and vertical directions, and $\|\cdot\|$ indicates the Euclidean norm. 
\par

Although the Tikhonov regularization successfully improves the image quality in astronomy (e.g. Kuramochi et al. \yearcite{2018ApJ...858...56K}\footnote{They call this term total squared variation.}), it is not necessarily optimal for galaxy images. 
This is because the Tikhonov regularization linearly calculates the image gradients despite the fact that galaxies have a super linear profile such as a S{\'e}rsic profile. 
As a natural consequence, the image gradients in logarithmic space can be more appropriate for the regularization. 
Hence, we suggest a Tikhonov regularization term on magnitude-image domain. 
One problem in the use of magnitude is a divergence when pixel values are zero or close to zero. 
To avoid this problem, we apply asinh magnitudes suggested in \citet{1999AJ....118.1406L}. 
\begin{eqnarray}
  \label{eq:amag}
    \mu(x) = -a \left[\sinh^{-1}\frac{x}{2b} + \ln{b} \right], 
\end{eqnarray}
where $a$ = $-2.5 \log e$ = 1.08574, and {\it b} is an arbitrary parameter to determine the linearity behaviour.
While $\mu(x)$ is linear in $x$ for $|x| \ll b$, it shows logarithmic behaviour for $|x| \gg b$.
Although it is natural to choose $b$ to be an flux uncertainty as discussed in \citet{1999AJ....118.1406L}, we discuss the optimal choice of $b$ in section \ref{hpar}.
\par

In summary, we propose the following formulation for a galaxy-image deblurring problem,
\begin{eqnarray}
  \label{eq:cost}
        \mathop{\arg\min}_\mathbf{X} \left[
          \; \frac{1}{2} \chi\left(\mathbf{X}\right)^2 + \lambda \|\mathbf{L\mu(X)}\|^2\;
          \right]
        s.t. \; \mathbf{X} > 0.
\end{eqnarray}

\subsection{Optimization}
\label{optimize}
The difficulty in solving the problem (\ref{eq:cost}) is the handling of the second term.
It has three functions to be calculated; the linear operator $\mathbf{L}$, Euclidean norm, and asinh magnitude.
To address this issue, we take a strategy that divides the variable $\mathbf{X}$ into two terms, $\mathbf{X_{flux}}$ and $\mathbf{X_{mag}}$, and updates them separately. 
Furthermore, the updates are conducted via a primal-dual splitting method \citep{Condat2013}, with which the calculations of linear operator and norm can be separated. 
This method has also another advantage that it can avoid an inner loop caused by a sub-problem, which is often appeared in the alternating direction method of multipliers \citep{ADMM_Boyd2011}.
In the following, we describe our strategy in detail. 
\par

Dividing the variable $\mathbf{X}$ into flux and magnitude terms, we modify the problem (\ref{eq:cost}) as the following. 
\begin{eqnarray}
  \label{eq:costmod}
  \mathop{\arg\min}_{ \mathbf{X_{flux}}, \mathbf{X_{mag}}} \; \frac{1}{2} \chi\left(\mathbf{X_{flux}}\right)^2 + \lambda \|\mathbf{LX_{mag}}\|^2, \nonumber \\
  s.t. \; \mathbf{X_{flux}} > 0\;
  and \; \mathbf{X_{mag}} = \mathbf{\mu(X_{flux})}.
\end{eqnarray}
In the modified problem, the non-linear function (\ref{eq:amag}), is removed from the objective function. 
\par

We solve the problem (\ref{eq:costmod}) via the primal-dual splitting method.
To adopt the primal-dual splitting to our problem, here we define some variables and functions.
We define the primal variable ${\bf \mathcal{X}}$, which have flux and magnitude parts, 
\begin{equation}
  \label{eq:primal}
        \mathbf{\mathcal{X}} = (\mathbf{X_{flux}} \; \mathbf{X_{mag}})^{\top}.
\end{equation}
To match the dimension, we also modify the observed image, dual variable, PSF matrix, and linear differential operator as,
${\bf \mathcal{Y}} = ({\bf Y} \; {\bf 0})^{\top}$,
${\bf \mathcal{V}} = ({\bf 0} \; {\bf V})^{\top}$,
${\bf \mathcal{A}} = ({\bf A} \; {\bf 0})$,
and ${\bf \mathcal{L}} = ({\bf 0} \; {\bf L})$. 
Using these definitions, we also introduce data fidelity term {\it F}, regularization term {\it H}, and constraints {\it G}. 
\begin{eqnarray}
  &&F(\mathbf{\mathcal{X}}) = \frac{1}{2} \chi^2\left(\mathbf{X_{flux}}\right),
  \label{eq:F}
  \\
  &&H(\mathbf{\mathcal{LX}}) = \lambda \|\mathbf{{LX}_{mag}}\|^2,
  \\
  &&G(\mathbf{\mathcal{X}}) = G_{1}(\mathbf{\mathcal{X}}) + G_{2}(\mathbf{\mathcal{X}}),
  \\
  &&G_{1}(\mathbf{\mathcal{X}}) = 
  \left\{
  \begin{array}{ll}
    \infty & \mathbf{X_{flux}} < 0 \\
    0 & {\rm else},\\
  \end{array}
  \right.\\
  &&G_{2}(\mathbf{\mathcal{X}}) = 
  \left\{
  \begin{array}{ll}
    \infty & \mathbf{X_{mag}} \ne \mu(\mathbf{X_{flux}})\\
    0 & {\rm else}.\\
  \end{array}
  \right.
\end{eqnarray}
With the above definitions, the problem(\ref{eq:costmod}) can be expressed as the following.
\begin{equation}
  \mathop{\arg\min}_{{\bf \mathcal{X}}}\;
  F(\mathbf{\mathcal{X}})
  + G(\mathbf{\mathcal{X}})
  + H(\mathbf{\mathcal{LX}}).
\end{equation}
This minimization problem is equivalent to a problem finding the saddle point of a Lagrangian,
\begin{equation}
  \label{eq:lagrangian}
  {\arg} \mathop{\min}_{\mathbf{\mathcal{X}}} \mathop{\max}_{{\mathbf{\mathcal{V}}}}\;
  \left[
  F(\mathbf{\mathcal{X}})
  + G(\mathbf{\mathcal{X}})
  - H^{*}(\mathbf{\mathcal{V}})
  + \left<\mathbf{\mathcal{LX}}, \mathbf{\mathcal{V}}\right>
  \right],
\end{equation}
where $\mathbf{\mathcal{V}}$ is the dual variable of $\mathbf{\mathcal{X}}$, and $H^{*}$ is the convex conjugate of $H$.
As can be seen, the linear operator $\mathcal{L}$ and the function $\mathcal{H}$  (corresponds to Euclidean norm) are separated in equation (\ref{eq:lagrangian}), and they can be easily calculated. This is the greatest advantage of the primal-dual splitting method. 
\par

The primal and dual variables can be updated as the following. 
\begin{eqnarray}
  \label{eq:pds}
  &&{\bf \tilde{\mathcal{X}}}_{n+1} = {\rm prox}_{\tau G} \left( {\bf \mathcal{X}}_{n} - \tau \left(\nabla F({\bf \mathcal{X}}_{n}) + {\bf\mathcal{L}}^{*}{\bf \mathcal{V}}_{n}\right)\right),\\
  &&{\bf \tilde{\mathcal{V}}}_{n+1} = {\rm prox}_{\sigma H^{*}}\left( {\bf \mathcal{V}}_{n} + \sigma {\bf\mathcal{L}}(2 \tilde{{\bf\mathcal{X}}}_{n+1} - {\bf\mathcal{X}}_{n})\right),\\
  &&({\mathcal{X}}_{n+1}, {\mathcal{V}}_{n+1}) = \rho_{n} (\tilde{{\mathcal{X}}}_{n+1}, \tilde{{\mathcal{V}}}_{n+1}) + (1 - \rho_{n}) ({\mathcal{X}}_{n}, {\mathcal{V}}_{n}), 
\end{eqnarray}
where $\sigma$, $\tau$, and $\rho_n$ are the update parameters described later, and prox indicates the proximal operator calculated as follows. 
%
The proximal operator with $G_{1}$ calculates the positivity constraint,
\begin{equation}
  \label{eq:proxg1}
  {\rm prox}_{\tau G_{1}}\left({\bf \mathcal{X}}\right) = \max\left(0, {\bf \mathcal{X}}\right).
\end{equation}
%
The proximal operator with $G_{2}$ calculates the consistency between $\mathbf{X_{flux}}$ and $\mathbf{X_{mag}}$, 
\begin{eqnarray}
  \label{eq:proxg2}
        &&{\rm prox}_{\tau G_{2}}\left(\mathbf{\mathcal{X}}\right)
        = \left(\mathbf{\check{X}_{flux, n}} \;\; \mathbf{\check{X}_{mag, n}}\right)^{\top}, \\
        &&\mathbf{\check{X}_{flux,n}} = \frac{\mathbf{X_{flux,n}} + \mathbf{w_{mag}}\mu^{-1}\left(\mathbf{X_{mag,n}}\right)}{1 + \mathbf{w_{mag}}},\\
        &&\mathbf{\check{X}_{mag,n}} = \mu\left(\mathbf{\check{X}_{flux,n}}\right),
\end{eqnarray}
where equation (22) indicates the weighted average of flux and magnitude with the weight vector $\mathbf{w_{mag}}$, calculated as follows.  
\begin{eqnarray}
  \label{eq:wmag}
  \mathbf{w_{mag}}
  &=& \frac{1}{(\mathbf{dX_{flux,n}} / \mathbf{dX_{mag,n}})^2} \nonumber \\
  &=& \left(\frac{a}{\sqrt{ (\mathbf{X_{flux,{\rm n}}})^2 + (2b)^2}} \right)^2.
\end{eqnarray}
%
The proximal operation of Euclidean norm can be calculated as follows. 
\begin{eqnarray}
  \label{eq:proxl2}
  {\rm prox}_{\sigma, H^{*}}\left({\bf \mathcal{X}}\right)
  &=& {\bf \mathcal{X}} - \sigma \, {\rm prox}_{\frac{1}{\sigma} H} \left({\bf\mathcal{X}}/\sigma\right) \nonumber \\
  &=& {\bf \mathcal{X}} - \sigma \frac{\left({\bf \mathcal{X}}/\sigma\right)\sigma}{2\lambda+\sigma} \nonumber\\
  &=& \frac{2\lambda}{2\lambda + \sigma} {\bf \mathcal{X}} \nonumber\\
  &=& \alpha {\bf \mathcal{X}},
\end{eqnarray}
where we used Moreau's identity (e.g. \cite{Bauschke11}) in the first equality and defined $\alpha {\rm =} 2\lambda/(2\lambda + \sigma)$ in the last equality. 
\par

Initial estimation of $\mathbf{X_{flux,0}}$ is arbitrary but we choose the observed image $\mathbf{Y}$.
We confirmed that images with zero or random values lead to the same restoration results although more iterations are required for convergence. It indicates that our method is not sensitive to the initial estimation. 
\par

%
There are three tuning parameters in the primal-dual splitting, $\tau$, $\sigma$, and $\rho_{n}$. 
All of these parameters are mainly related to convergence speed, and not sensitive to the resultant image quality. 
The parameters $\tau$ and $\sigma$ are related to updates of the primal and dual variables, and should be $0 < \tau < 2/\beta$ and $\sigma \le \left(\frac{1}{\tau} - \frac{\beta}{2} \right)/\|L\|^2$, where $\beta$ is the Lipschitz constant of the equation (\ref{eq:F}), 
\begin{equation}
  \label{eq:beta}
  \beta = \|\mathbf{A^{\top}A}\| / \min(\mathbf{\Sigma}).
\end{equation}
Although the choice of $\tau$ and $\sigma$ is arbitrary, we define them as the following.
\begin{equation}
  \label{eq:tau}
  \tau = 1/\beta,
\end{equation}
\begin{eqnarray}
  \label{eq:sig}
  \sigma &=& (\frac{1}{\tau} - \frac{\beta}{2}) / \|L\|^2 \nonumber \\
  &=& \frac{\beta}{2 \|L\|^2}.
\end{eqnarray}
With the above definition, these parameters always satisfy their constraint.
The parameter $\rho_{n}$ is an update rate and can be modified during the iteration.
The optimal choice of $\rho_{n}$ may accelerate the convergence, making use of the previous iteration. 
Nonetheless, we simply adopt $\rho_{n}$ of one, that is, we do not use the information from the previous update. 
\par

With the above setting, we can solve the problem (\ref{eq:lagrangian}).
Nonetheless, the convergence might be very slow,  depending on the parameters $\sigma$ and $\lambda$.
This is because, if $\lambda \gg \sigma$, 
$\alpha$ in equation (\ref{eq:proxl2}) converges to 1 and not sensitive to the regularization parameter $\lambda$.
This means that many iterations are required to reflect the regularization into the restored image. 
To avoid this situation, we scale the image and the variance beforehand. 
The scaling parameter is as follows.
\begin{eqnarray}
  \label{eq:scale}
  {\rm scale} = \sqrt{ \frac{\|\mathbf{A^{\top}A}\|}{2\lambda\min(\mathbf{\Sigma}) \|L\|^2}   }.
\end{eqnarray}
This scaling leads to $\sigma=\lambda$ and $\alpha=0.66$.
Although $\alpha$ is independent of $\lambda$, the primal variable strongly depends on $\lambda$ because the scaling factor is proportional to $1/\sqrt{\lambda}$. 
This scaling as a pre-processing significantly accelerates the convergence of our problem. 
\par

\subsection{Comparison methods}
The main objective of this paper is to investigate the feasibility of Tikhonov regularization on magnitude domain. 
For this purpose, we compare our method with a widely applied RL method and a method with conventional Tikhonov regularization on flux domain. We explain these methods in the following. 

\subsubsection{Richardson-Lucy algorithm}
The RL method is a maximum-likelihood method assuming pixel values to follow a Poisson distribution.
A given initial estimated image is iteratively updated to increase the likelihood and the positivity constraint is automatically satisfied. 
The update at the {\it n}th iteration is conducted as follows.
\begin{eqnarray}
  \mathbf{X}_{n+1} &=& \mathbf{X}_{n} * \mathbf{A}^{\top} \mathbf{Z}_{n+1}, \\
  \mathbf{Z}_{n+1} &=& \mathbf{Y} / (\mathbf{A} \mathbf{X}_{n}), \nonumber
\end{eqnarray}
where $*$ and $/$ stand for multiplication and division of the same elements. 
Since the iteration should monotonically increase the likelihood, the infinite number of iterations would lead to a maximum likelihood solution. 
However, the iteration should stop at some time to avoid the divergence, i.e. the noise amplification. 
The stopping criterion is somewhat arbitrary and can be regarded as a hyper parameter. 
We choose the number of iterations to maximize the restored image quality although it is impossible to determine it in reality because we do not have the ground-truth. 
This choice leads to an overestimation of the RL performance, but our method still outperforms the RL method as shown in later. 

\subsubsection{Tikhonov regularization on flux domain}
Conventional Tikhonov regularization methods adopt a regularization on flux domain. 
Hence, the objective function is very similar to ours and the only difference is asinh magnitude in equation (\ref{eq:cost}). 
Hereafter we simply refer this method as the Tikhonov method. 
For a fare comparison, we perform the same optimization method as the proposed method, the primal-dual splitting method.
\par

%
\section{Evaluation}
\label{sec:evaluation}
We evaluate the capability of the proposed method using simulation and observation images. 
In the following, we describe our evaluation methods and quantitative metrics. 

\subsection{Simulation}
\label{sec:sim}
Simulation studies are suitable for a principle demonstration of our method because we have ground-truth images and perfect PSFs.  
The former can be directly compared with the reconstructed images, and owing to the latter we can also directly evaluate our method without uncertainty in PSFs. 
\par

We retrieved one spiral and one elliptical galaxies from the {\it Subaru} HSC-SSP DR2 \citep{2019PASJ...71..114A} observed with the {\it i} band. 
Hereafter we call them Spiral-1 and Elliptical-1, respectively. 
The use of different morphologies is suitable for performance verification. 
We visually chose the galaxies using the HSC map\footnote{https://hsc-release.mtk.nao.ac.jp/doc/index.php/tools-2/} and show their coordinates and redshifts in the upper side of Table \ref{tab:target}.
The galaxies have moderate redshifts, so that we can ignore the PSF blurring and avoid saturation. 
We resampled and shrank the images to images with 64$\times$64 pixels.
The magnification factors of the pixel scales were 4.2 and 2.1 for Spiral-1 and Elliptical-1, respectively\footnote{It corresponds to pixel scales of 0.7 and 0.35 arcsec/pixel while the original is 0.168 arcsec/pixel.}.
Figures \ref{fig:spi1}a and \ref{fig:ell1}a show their resampled images. 
\par

The images were blurred with a realistic PSF. 
While Gaussian functions are often applied as a PSF, they are clearly distinct from realistic ones. 
We used a more realistic PSF. We combined the PSFs for Spiral-1 and Elliptical-1, which were provided by the HSC project\footnote{https://hsc-release.mtk.nao.ac.jp/psf/pdr2/}.
The FWHM of the combined PSF was $\sim$3 pixel. 
Because the images were shrank as mentioned in the previous paragraph, the convolution with the combined PSFs significantly blurred the images. 

We added noises to the blurred images. 
To represent photon noises from both the background and objects, we applied Gaussian and Poisson distributions, respectively.
We note that the sky background had been subtracted from the images and were assumed to be uniform across the images, and the sky noise was able to be expressed as a Gaussian distribution. 
We set the standard deviation of the Gaussian noise to 1.0\% of the peak signal, whereas we produced two Poisson noises: the standard deviations of 1.0\% and 5.0\% of the peak signal. 
Hereafter the noise levels are referred to as 1.0\% and 5.0\% despite the additional 1.0\% Gaussian noise. 
The procedure of the noise production is the same for both Spiral-1 and Elliptical-1.
Panels (b) and (f) in figures \ref{fig:spi1} and \ref{fig:ell1} show the blurred Spiral-1 and Elliptical-1 images with 1.0\% and 5.0\% noises, respectively.
The S/N at the centres are about 70 and 20 for both morphologies, respectively. 
\par
We adopted our deblurring method to the images.
We set the hyper parameter $\lambda$ to 2.0 for Spiral-1 while 16.0 and 4.0 for Elliptical-1 with 1.0\% and 5.0\% noises, respectively. 
The other parameter {\it b} was set to 1.0$\bar{\sigma}_{y}$, where $\bar{\sigma}_{y}$ is the median flux uncertainty in the blurred image.
The number of iterations was 1000 unless otherwise stated. 
We discuss the appropriate parameter values in section \ref{hpar}.

\subsection{Application on real images}
\label{appreal}
To evaluate our method with more realistic situations, we applied it on real observation images. 
In contrast to the simulation, observation images could have unknown and/or systematic errors so that PSF deconvolution becomes a more difficult problem. 
We chose one spiral and one elliptical galaxies from the HSC-SSP DR2 as listed in the lower side of Table \ref{tab:target}.
Hereafter we call them Spiral-2 and Elliptical-2. 
They have an appropriate size to fit to 64$\times$64 pixel images as shown in panels (b) and (f) of figures \ref{fig:spi2} and \ref{fig:ell2}. 
We obtained the images, the variances, and the PSFs of each galaxy from the public data release site\footnote{https://hsc-release.mtk.nao.ac.jp/doc/index.php/tools-2/}.
We retrieved the images from both HSC-DUD and HSC-Wide surveys to obtain deep and shallow images, respectively.
The S/N at the centres of Spiral-2 and Elliptical-2 were 409 and 178 for HSC-DUD, and 105 and 53 for HSC-Wide, respectively. 
We used only {\it i-}band images since other band images have no reference {\it HST} images as described later.
The typical seeing at the {\it i-}band was $\sim$ 0.6 arcsec \citep{2019PASJ...71..114A}. 
We did not use mask information because the images were produced with a sufficient number of observations and no noticeable artefacts were found. 
\par
These observed images were deblurred with our method. 
We set $\lambda$ to 2.0 and {\it b} to 1.0$\bar{\sigma}_{y}$ for both Spiral-2 and Elliptical-2. 
We discuss the appropriate parameter values in section \ref{hpar}. 

Since we did not have ground-truth images, we instead used images taken with the {\it HST}.
The {\it HST} has the F814W filter, whose wavelength coverage is similar to the {\it i}-band, so that the images can be used as a proxy of the ground-truth. 
The pixel scale was 0.05 arcsec, which were resampled to 0.168 arcsec to match the pixel scale of the {\it Subaru}/HSC. 
To avoid uncertainties in the flux calibration and sky subtraction, we scaled the images and added offsets to them to match the restored HSC {\it i}-band images. 
The {\it HST} images were obtained from the Hubble Legacy Archive \citep{2006ASPC..351..406J,2008ASPC..394..481W,2016AJ....151..134W}. 

\subsection{Metrics}
For quantitative image evaluation, we applied two widely applied metrics, the peak signal to noise ratio (PSNR) and structural similarity (SSIM; \cite{2004ITIP...13..600W}). 
The PSNR of image $\boldsymbol{X}$ is defined as
\begin{eqnarray}
  \label{eq:psnr}
  \mbox{PSNR}(\boldsymbol{X,X_{ref}}){\rm [dB]} &=& 20\log_{10}\left(\frac{\mbox{MAX}(\boldsymbol{X_{ref}})}{\sqrt{\mbox{MSE}}}\right),\\
  \mbox{MSE} &=& \frac{1}{N}\sum_i\left(x_i - x_{ref,i}\right)^2,
\end{eqnarray}
where $N$ is the number of elements, and the subscript $ref$ indicates the ground-truth image.
By definition, images similar to the ground-truth have a high PSNR.
\par
The definition of SSIM is the following. 
\begin{eqnarray}
  &&\mbox{SSIM}(\boldsymbol{X,X_{ref}}) \nonumber \\
  &&= l(\boldsymbol{X,X_{ref}}) \cdot c(\boldsymbol{X,X_{ref}}) \cdot s(\boldsymbol{X,X_{ref}}), 
\end{eqnarray}
\begin{eqnarray}
  l(\boldsymbol{X,X_{ref}}) &=& \frac{2\mu_x \, \mu_{ref} + C_1}{\mu_x^2 + \mu_{ref}^2 + C_1}, \\ 
  c(\boldsymbol{X,X_{ref}}) &=& \frac{2\sigma_x \, \sigma_{ref} + C_2}{\sigma_x^2 + \sigma_{ref}^2 + C_2}, \\ 
  s(\boldsymbol{X,X_{ref}}) &=& \frac{\sigma_{x,ref} + C_3}{\sigma_x \, \sigma_{ref} + C_3},\\ 
  C_1 &=& \left(K_1\,L\right)^2,\\
  C_2 &=& \left(K_2\,L\right)^2,\\
  C_3 &=& C_2 / 2,
\end{eqnarray}
where $\mu$ is the average value, $\sigma_x$ is the standard deviation, $\sigma_{x,ref}$ is the covariance, $L$ is the dynamic range of the image, and $K_1$ and $K_2$ are arbitrary small values, where we adopted 0.001, following \citet{2004ITIP...13..600W}. 
The terms $l$, $c$, and $s$ correspond to comparisons between the image and reference in terms of the luminance, contrast, and structure, respectively.
When the image and reference are similar, these terms should be close to one, and the SSIM should also be close to one. 
\par

Since our simulation has a reference image for each galaxy, these metrics can be appropriate for our evaluation. 
On the other hand, the {\it HST} images are not perfect ground-truth ones for the observational evaluation.
For example, when the restored HSC images have better image quality, the {\it HST} images do not work as the ground-truth.
Hence, a care must be taken for the evaluation results.


\begin{threeparttable}
  \caption{Target galaxies used for the evaluation with simulation and observation.   
    \label{tab:target}}
  \begin{tabular}{cccc}
    \hline
    Name & R.A. & Dec. & z \\ 
    \hline\hline
    \multicolumn{4}{c}{Simulation}\\
    \hline
    Spiral-1 & 150.746077 & 2.342989 & 0.044 \\ 
    Elliptical-1 & 150.8170172 & 3.4772409 & 0.106\\
    \hline
    \multicolumn{4}{c}{Observation}\\
    \hline
    Spiral-2 & 150.1728195 & 2.6090672 & 0.265 \\
    Elliptical-2 & 150.687837 & 2.61864322 & 0.860\tnote{a}\\
    \hline
  \end{tabular}
  \begin{tablenotes}
  \item[a] photometric redshift from \citet{2018PASJ...70S...9T}.
  \end{tablenotes}
  \vspace{5mm}
\end{threeparttable}

\section{Results \& Discussion}

\subsection{Image quality}
\label{quality}
\begin{figure*}
  \begin{minipage}{0.19 \hsize}
    \center{\includegraphics[width=80pt]{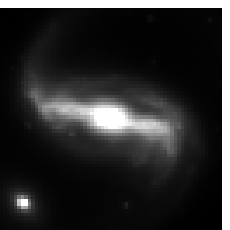} \\
      (a)Original}
  \end{minipage}
  \begin{minipage}{0.21 \hsize}
    \center{\includegraphics[width=80pt]{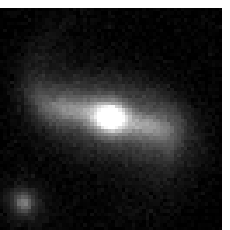} \\
      (b)Blurred($n=1\%$)}
  \end{minipage}
  \begin{minipage}{0.19 \hsize}
    \center{\includegraphics[width=80pt]{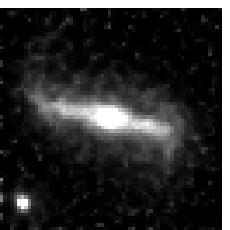} \\
      (c)RL}
  \end{minipage}
  \begin{minipage}{0.19 \hsize}
    \center{\includegraphics[width=80pt]{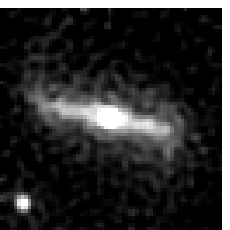} \\
      (d)Tikhonov}
  \end{minipage}
  \begin{minipage}{0.19 \hsize}
    \center{\includegraphics[width=80pt]{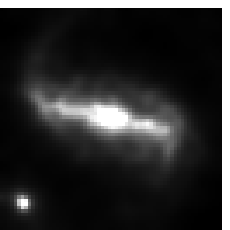} \\
      (e)This work}
  \end{minipage}
  \\
  \vspace{0.5 cm}
  
  \begin{minipage}{0.19 \hsize}
    \hspace{0.19 \hsize}
  \end{minipage}
  \begin{minipage}{0.21 \hsize}
    \center{\includegraphics[width=80pt]{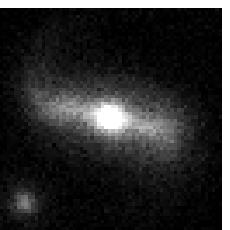} \\
      (f)Blurred($n=5\%$)}
  \end{minipage}
  \begin{minipage}{0.19 \hsize}
    \center{\includegraphics[width=80pt]{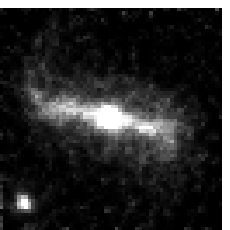} \\
      (g)RL}
  \end{minipage}
  \begin{minipage}{0.19 \hsize}
    \center{\includegraphics[width=80pt]{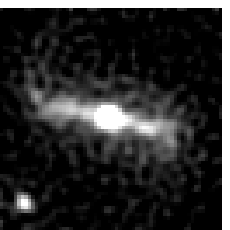} \\
      (h)Tikhonov}
  \end{minipage}
  \begin{minipage}{0.19 \hsize}
    \center{\includegraphics[width=80pt]{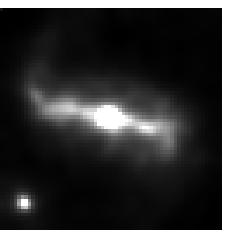} \\
      (i)This work}
  \end{minipage}
  \\
  \vspace{0.2cm}
  \caption{The comparison of restored Spiral-1 images in the simulation. 
    \label{fig:spi1}}
\end{figure*}

\begin{figure*}
  \begin{minipage}{0.19 \hsize}
    \center{\includegraphics[width=80pt]{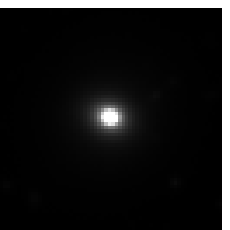}\\
      (a)Original}
  \end{minipage}
  \begin{minipage}{0.21 \hsize}
    \center{\includegraphics[width=80pt]{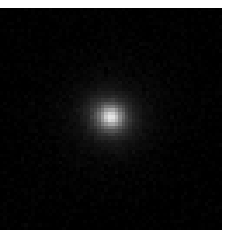}\\
      (b)Blurred($n=1\%$)}
  \end{minipage}
  \begin{minipage}{0.19 \hsize}
    \center{\includegraphics[width=80pt]{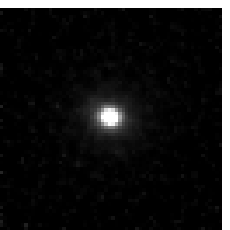}\\
      (c)RL}
  \end{minipage}
  \begin{minipage}{0.19 \hsize}
    \center{\includegraphics[width=80pt]{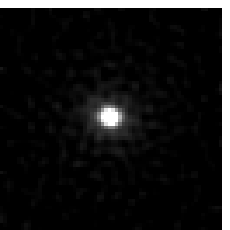}\\
      (d)Tikhonov}
  \end{minipage}
  \begin{minipage}{0.19 \hsize}
    \center{\includegraphics[width=80pt]{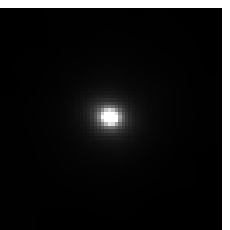}\\
      (e)This work}
  \end{minipage}
  \\
  \vspace{0.5 cm}
  
  \begin{minipage}{0.19 \hsize}
    \hspace{0.19 \hsize}
  \end{minipage}
  \begin{minipage}{0.21 \hsize}
    \center{\includegraphics[width=80pt]{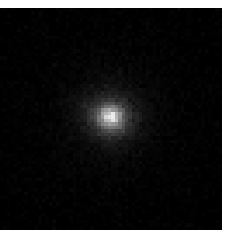}\\
      (f)Blurred($n=5\%$)}
  \end{minipage}
  \begin{minipage}{0.19 \hsize}
    \center{\includegraphics[width=80pt]{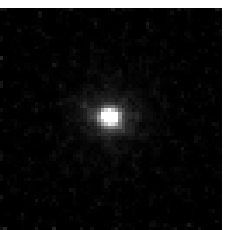}\\
      (g)RL}
  \end{minipage}
  \begin{minipage}{0.19 \hsize}
    \center{\includegraphics[width=80pt]{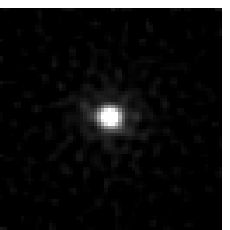}\\
      (h)Tikhonov}
  \end{minipage}
  \begin{minipage}{0.19 \hsize}
    \center{\includegraphics[width=80pt]{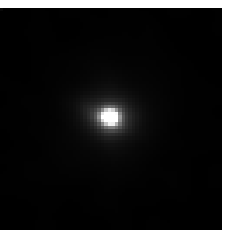}\\
      (i)This work}
  \end{minipage}
  \\
  \vspace{0.2cm}
  \caption{The same as figure \ref{fig:spi1} but for Elliptical-1. 
    \label{fig:ell1}}
\end{figure*}

Figures \ref{fig:spi1} and \ref{fig:ell1} show the image-restoration results for Spiral-1 and Elliptical-1, respectively. 
While all the three methods (RL, Tikhonov, and this work) successfully restore the blurred and noisy images (panels b and f), our method shows the best performance. 
Remarkable distinction can be seen especially in noisier images (bottom panels). 
We show the quantitative comparison in table \ref{tab:evasim}. 
In the following, we discuss the image qualities in detail. 

\par
Reconstructed images with the RL method (panels c and g) show higher noise despite the higher spatial resolution than the blurred ones (panels b and f). 
This is because the iteration of the RL method amplifies the noise while improving the resolution. 
Owing to the trade-off, the number of iterations should be optimized. 
We determined the number to maximize the PSNR.
While the best numbers of iterations were slightly different for PSNR and SSIM, we show the maximum values of them in table \ref{tab:evasim}. 
Whereas we can determine the best number of iterations in a simulation study, we could not in general, so that the use of the best numbers leads to unfair comparisons. 
Nonetheless, the restored images with our method still show higher image qualities than those from the RL method. 
For example, for the Spiral-1 and Elliptical-1 images with 1\% noises, the RL method provides the PSNRs of 40.39 and 43.98 dB whereas ours show 43.07 and 57.11 dB, respectively. 
These results strongly suggest that our method outperforms the RL method. 
We have to note that the image quality is affected by the number of iterations even in our method, which we discuss in subsection \ref{convergence}.
Since the increase of the number of iterations after exceeding 1000 affects only a few central pixels, which could not be seen visually, we show only images with the 1000 iterations in the figures \ref{fig:spi1} and \ref{fig:ell1}.

\par
Panels (d) and (h) in figures \ref{fig:spi1} and \ref{fig:ell1} show the Spiral-1 and Elliptical-1 images that have 1 and 5\% noises and are restored with the Tikhonov regularization. 
Compared with the results from the RL method, high frequency components caused by noises are reduced owing to the regularization, but the images are blurred. 
This results in lower PSNRs and SSIMs as shown in table \ref{tab:evasim}.
The low image qualities resulted from the regularization suggest that the assumed prior is not appropriate. 
Since both spiral and elliptical galaxies tend to have intensity profiles that exponentially change with the position, minimizing the gradients in linear space leads to blurring detailed structures, as mentioned in the introduction. 
In contrast, images restored with our method (panels e and i) are appropriately smoothed but remarkably reproduce the galaxy structures. 
The quantitative comparison in table \ref{tab:evasim} also shows that our method leads to the best qualities. 
These results suggest that the proposed regularization, a quadratic norm of the differential image in magnitude space (i.e. log space), is appropriate. 

\par
Since the elliptical galaxies are compact and the intensity profiles rapidly change, the difference of the restored images may not be obvious. 
To see the difference more clearly, we plot in figure \ref{fig:ellprf}a the horizontal profiles at the centre of the Elliptical-1 images, where the added noise level is 5.0\%. 
At the first look, any de-blurring method produces sharper profiles than the blurred one (grey).
Among them, our method (bright cyan) shows the most similar profile to the original (black). 

\par
\begin{figure*}
  \begin{minipage}{0.19 \hsize}
    \center{\includegraphics[width=80pt]{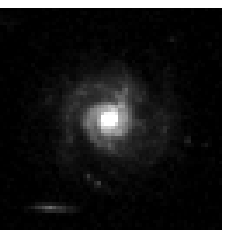}\\
      (a)HST}
  \end{minipage}
  \begin{minipage}{0.19 \hsize}
    \center{\includegraphics[width=80pt]{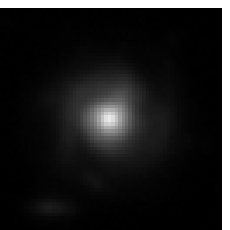}\\
      (b)HSC-DUD}
  \end{minipage}
  \begin{minipage}{0.19 \hsize}
    \center{\includegraphics[width=80pt]{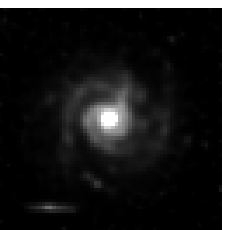}\\
      (c)RL}
  \end{minipage}
  \begin{minipage}{0.19 \hsize}
    \center{\includegraphics[width=80pt]{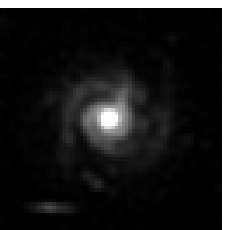}\\
      (d)Tikhonov}
  \end{minipage}
  \begin{minipage}{0.19 \hsize}
    \center{\includegraphics[width=80pt]{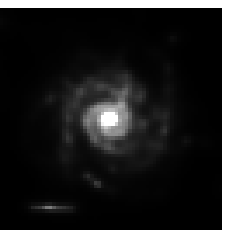}\\
      (e)This work}
  \end{minipage}
  \vspace{0.5 cm}
  
  \begin{minipage}{0.19 \hsize}
    \hspace{0.16\hsize}
  \end{minipage}
  \begin{minipage}{0.19 \hsize}
    \center{\includegraphics[width=80pt]{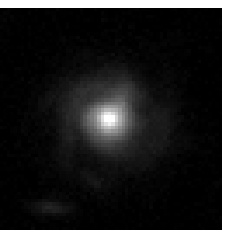}\\
      (f)HSC-Wide}
  \end{minipage}
  \begin{minipage}{0.19 \hsize}
    \center{\includegraphics[width=80pt]{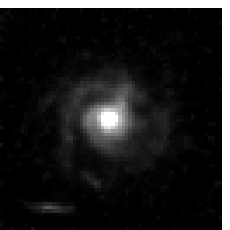}\\
      (g)RL}
  \end{minipage}
  \begin{minipage}{0.19 \hsize}
    \center{\includegraphics[width=80pt]{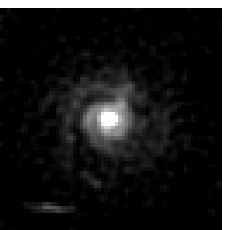}\\
      (h)Tikhonov}
  \end{minipage}
  \begin{minipage}{0.19 \hsize}
    \center{\includegraphics[width=80pt]{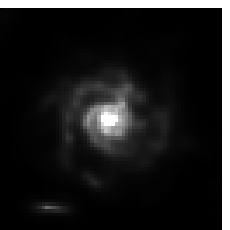}\\
      (i)This work}
  \end{minipage}
  \\
  \vspace{0.2cm}
  \caption{
    The comparison of restored Spiral-2 images. 
    \label{fig:spi2}}
\end{figure*}
\begin{figure*}
  \begin{minipage}{0.19 \hsize}
    \center{\includegraphics[width=80pt]{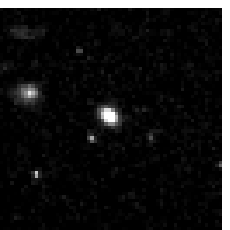}\\
      (a)HST}
  \end{minipage}
  \begin{minipage}{0.19 \hsize}
    \center{\includegraphics[width=80pt]{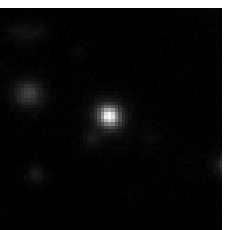}\\
      (b)HSC-DUD}
  \end{minipage}
  \begin{minipage}{0.19 \hsize}
    \center{\includegraphics[width=80pt]{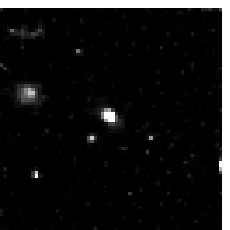}\\
      (c)RL}
  \end{minipage}
  \begin{minipage}{0.19 \hsize}
    \center{\includegraphics[width=80pt]{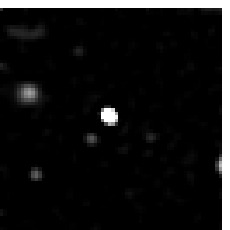}\\
      (d)Tikhonov}
  \end{minipage}
  \begin{minipage}{0.19 \hsize}
    \center{\includegraphics[width=80pt]{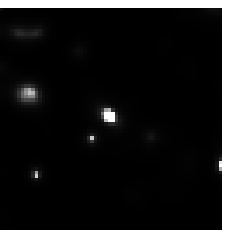}\\
      (e)This work}
  \end{minipage}
  \\
  \vspace{0.5 cm}
  
  \begin{minipage}{0.19 \hsize}
    \hspace{0.19\hsize}
  \end{minipage}
  \begin{minipage}{0.19 \hsize}
    \center{\includegraphics[width=80pt]{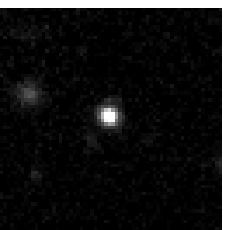}\\
      (f)HSC-Wide}
  \end{minipage}
  \begin{minipage}{0.19 \hsize}
    \center{\includegraphics[width=80pt]{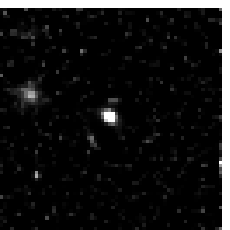}\\
      (g)RL}
  \end{minipage}
  \begin{minipage}{0.19 \hsize}
    \center{\includegraphics[width=80pt]{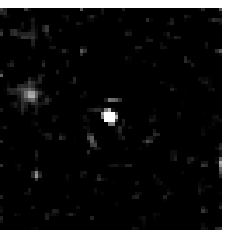}\\
      (h)Tikhonov}
  \end{minipage}
  \begin{minipage}{0.19 \hsize}
    \center{\includegraphics[width=80pt]{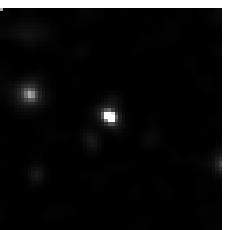}\\
      (i)This work}
  \end{minipage}
  \\
  \vspace{0.2cm}
  \caption{The same as figure \ref{fig:spi2}  but for Elliptical-2. 
    \label{fig:ell2}}
\end{figure*}

Figures \ref{fig:spi2} and \ref{fig:ell2} show the results on real images (Spiral-2 and Elliptical-2, respectively).
Similar to the simulation results, our method best restores the detailed structures in the galaxies. 
The quantitative evaluation in table \ref{tab:evaobs} also shows that our method basically produces the highest PSNRs and SSIMs. 
The only case our method leads to lower PSNR than the RL is for Elliptical-2 in HSC-DUD.
This is because the S/N exceeds 170 as mentioned in section \ref{appreal}. Although the RL method is sensitive to a noise, it can work for high S/N images. 
On the other hand, the lower S/N case, i.e. in the HSC-Wide, our method produces a higher PSNR than the RL.  
It indicates that our method can work even for low S/N images.
\par

\subsection{Convergence}
\label{convergence}
Here we discuss the optimal number of iterations for our method. 
It is generally difficult to determine the number of iterations. 
While increasing the number of iterations decreases the objective function, it directly leads to a high computational cost and does not necessarily improve the quality of restored images. 
Hence, the number of iterations should be optimized by balancing the computational costs with the resulting image quality.  
\par

\begin{figure*}
  \begin{minipage}{0.49 \hsize}
    \center{\includegraphics[width=0.8\hsize]{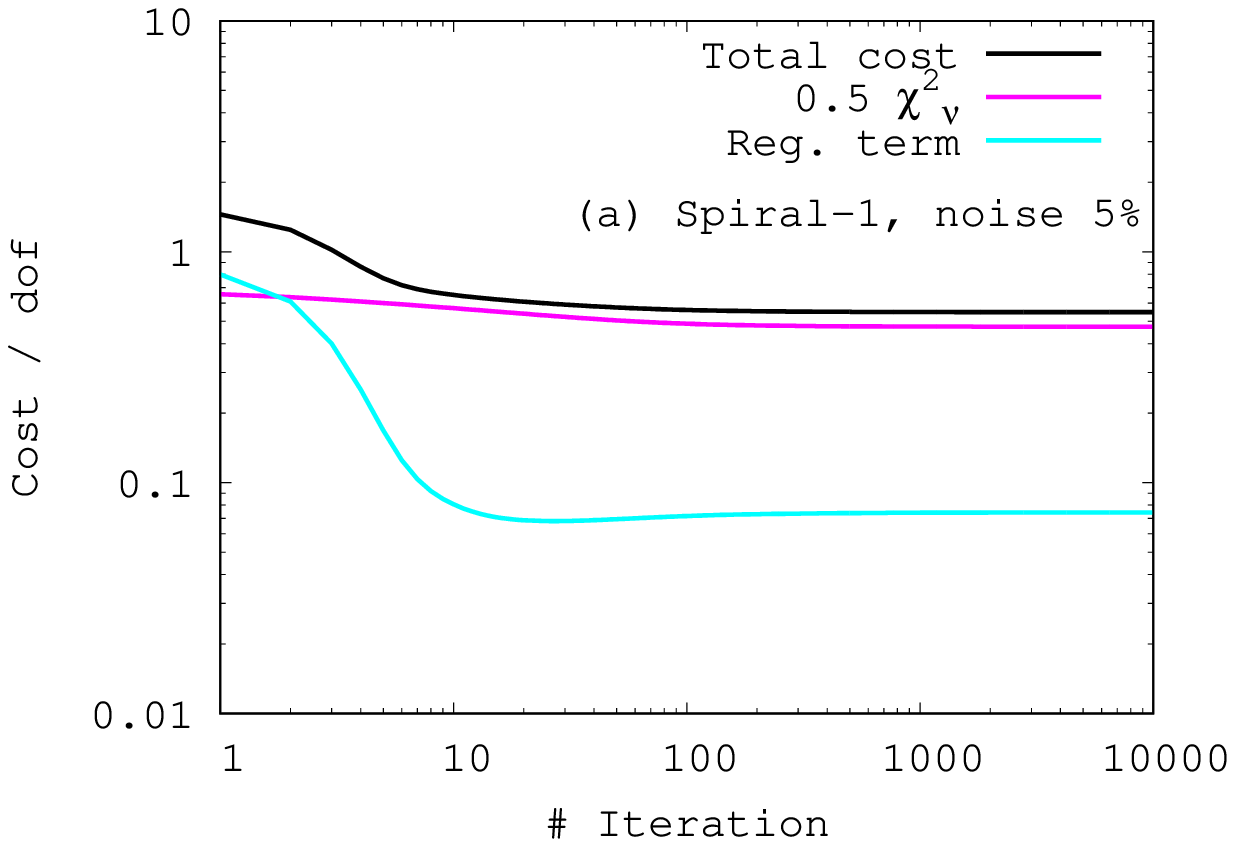}}
  \end{minipage}
  \begin{minipage}{0.49 \hsize}
    \center{\includegraphics[width=0.8\hsize]{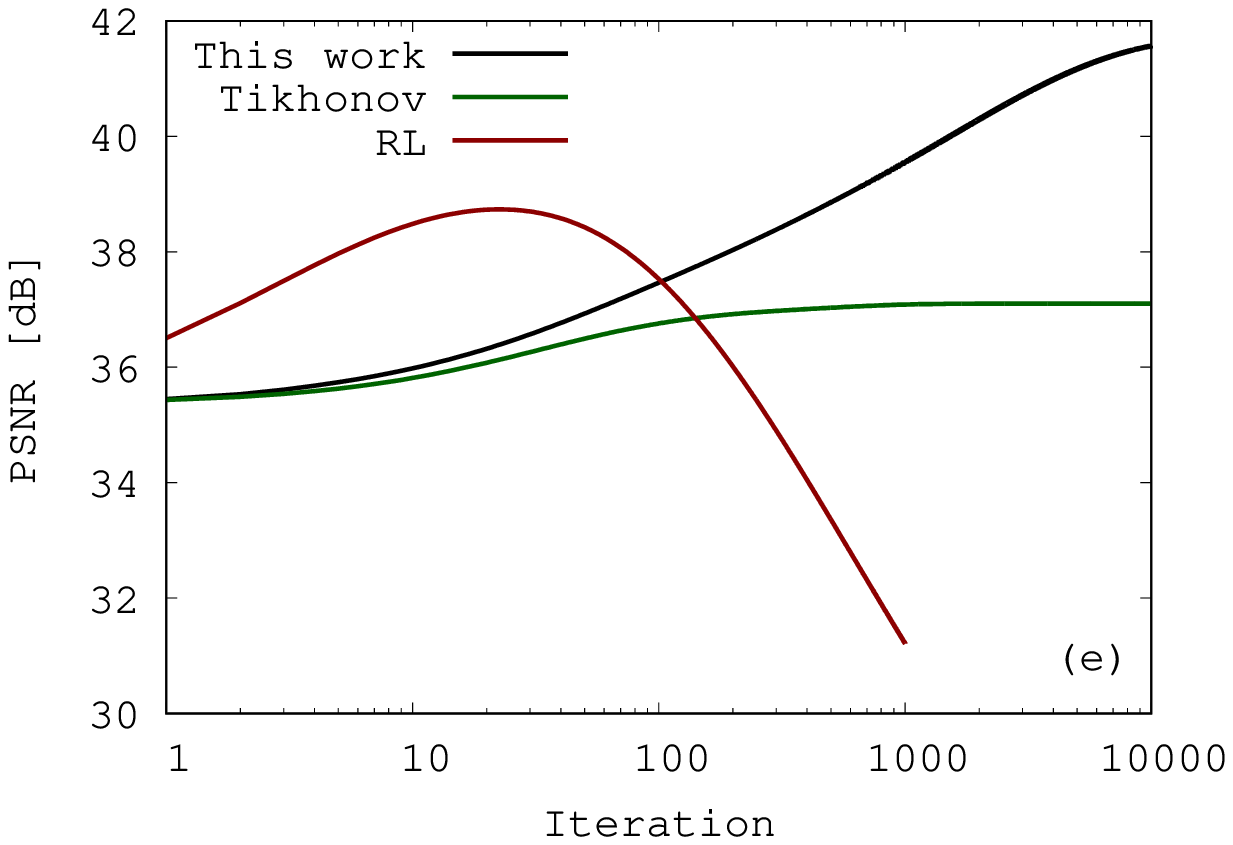}}
  \end{minipage}
  \\
  \vspace{0.5 cm}
  \begin{minipage}{0.49 \hsize}
    \center{\includegraphics[width=0.8\hsize]{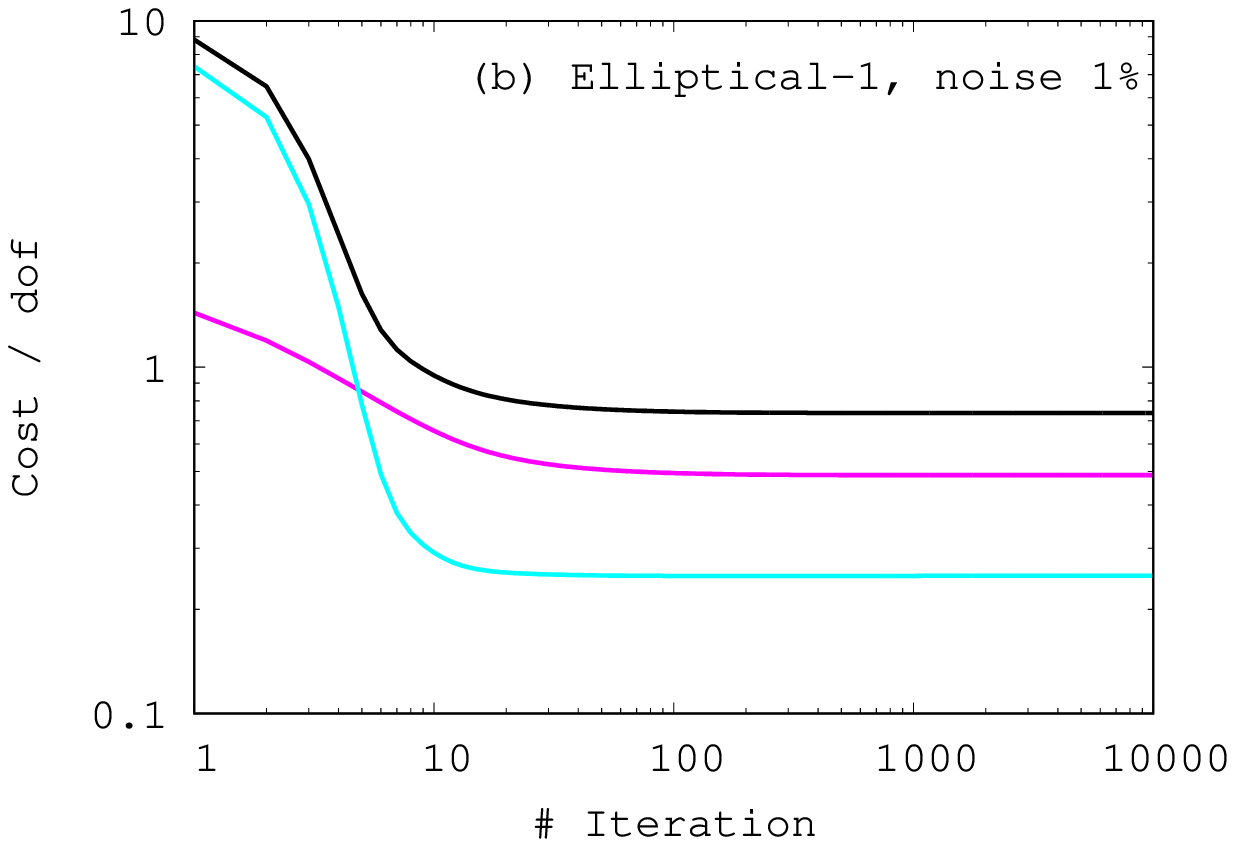}}
  \end{minipage}
  \begin{minipage}{0.49 \hsize}
    \center{\includegraphics[width=0.8\hsize]{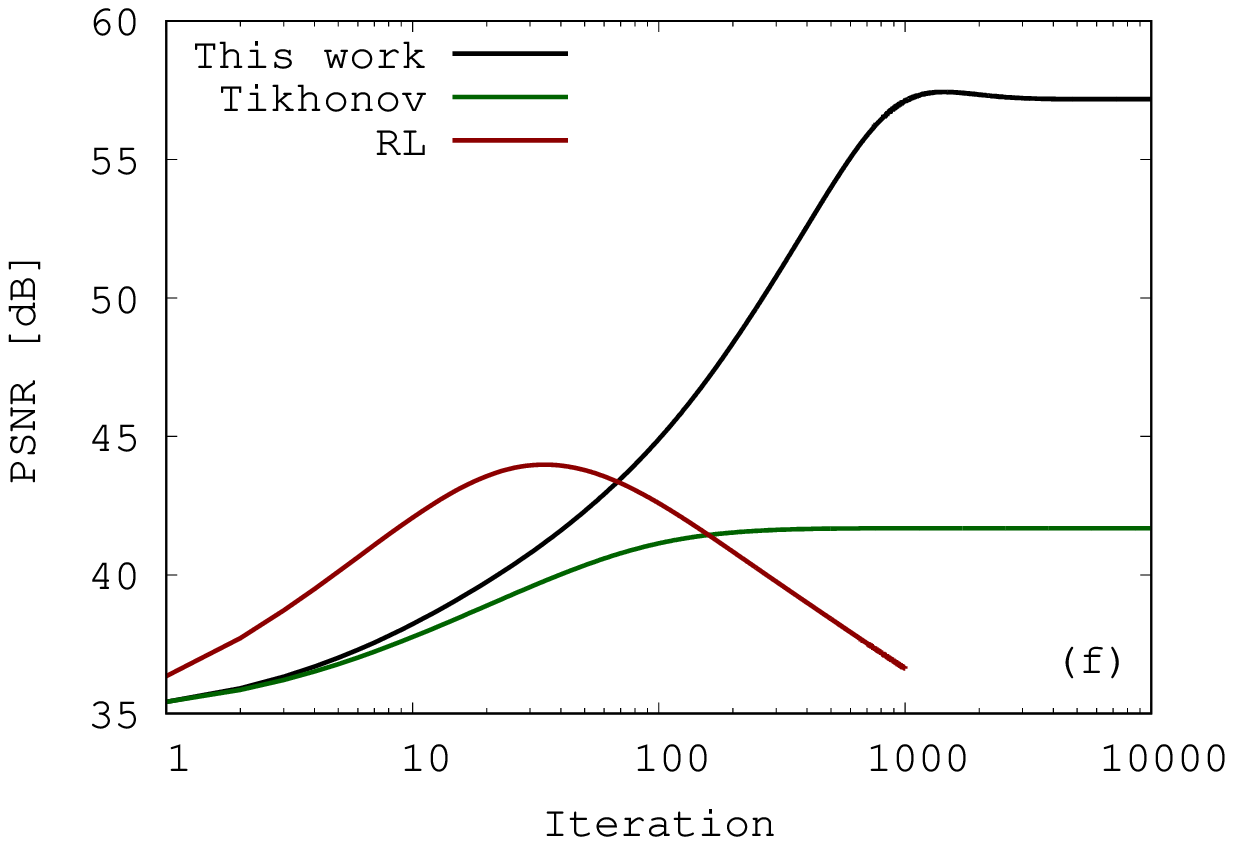}}
  \end{minipage}
  \\
  \vspace{0.5 cm}
  \begin{minipage}{0.49 \hsize}
    \center{\includegraphics[width=0.8\hsize]{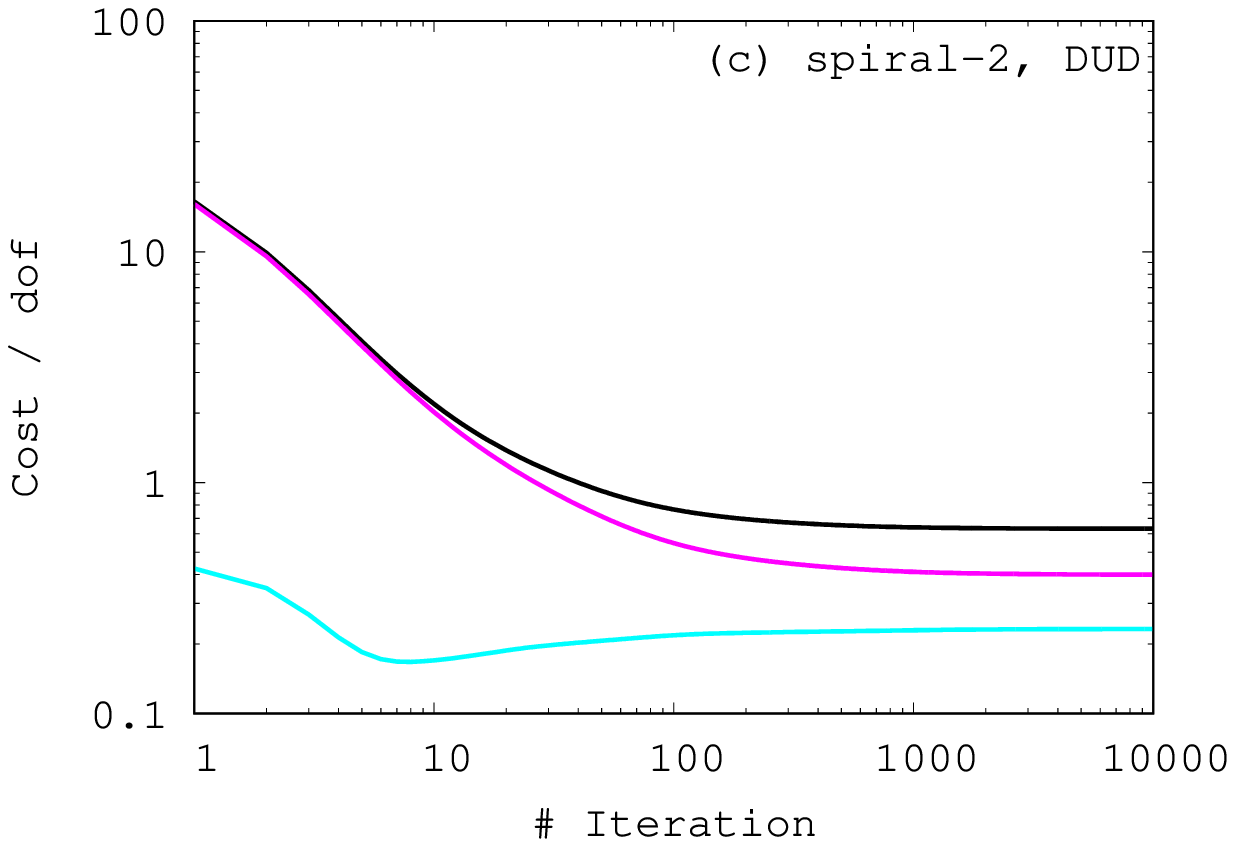}}
  \end{minipage}
  \begin{minipage}{0.49 \hsize}
    \center{\includegraphics[width=0.8\hsize]{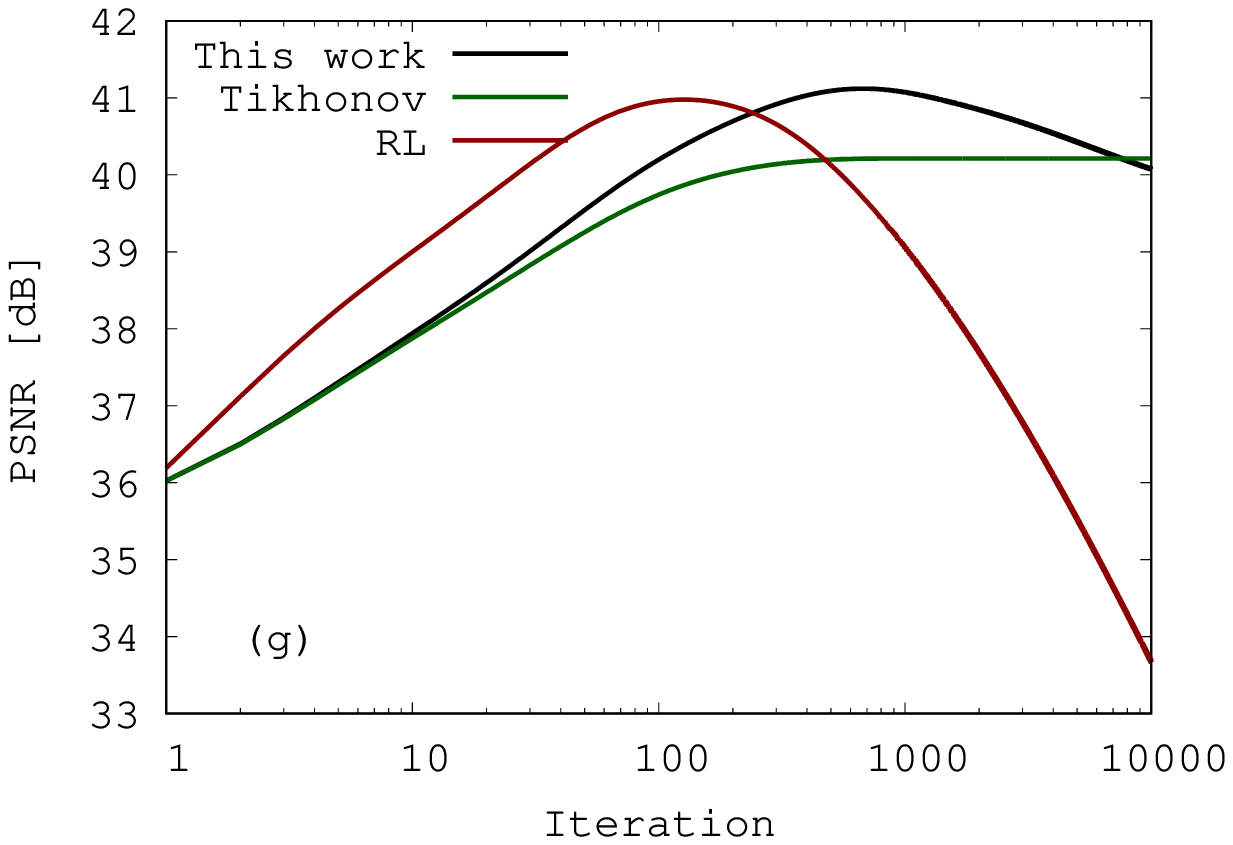}}
  \end{minipage}
  \\
  \vspace{0.5 cm}
  \begin{minipage}{0.49 \hsize}
  \center{\includegraphics[width=0.8\hsize]{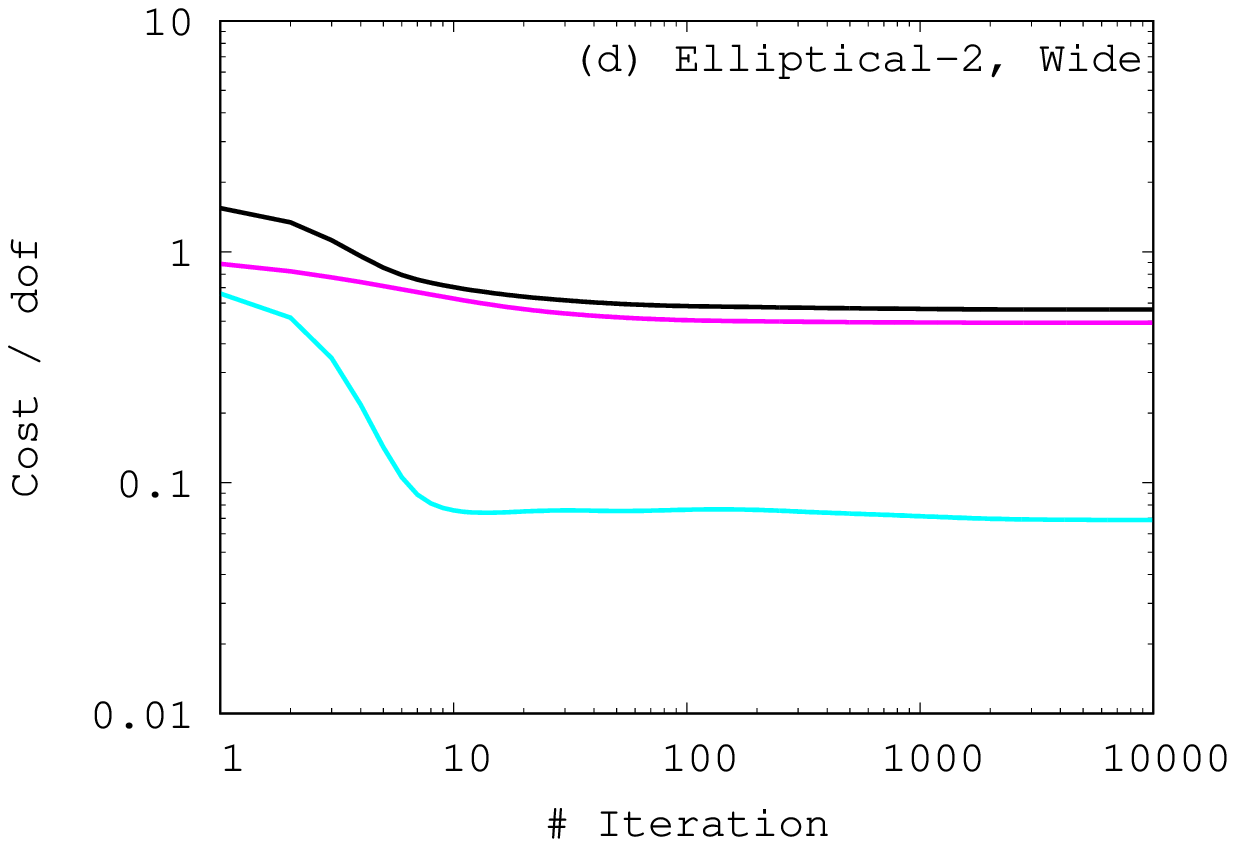}
    }
  \end{minipage}
  \begin{minipage}{0.49 \hsize}
    \center{\includegraphics[width=0.8\hsize]{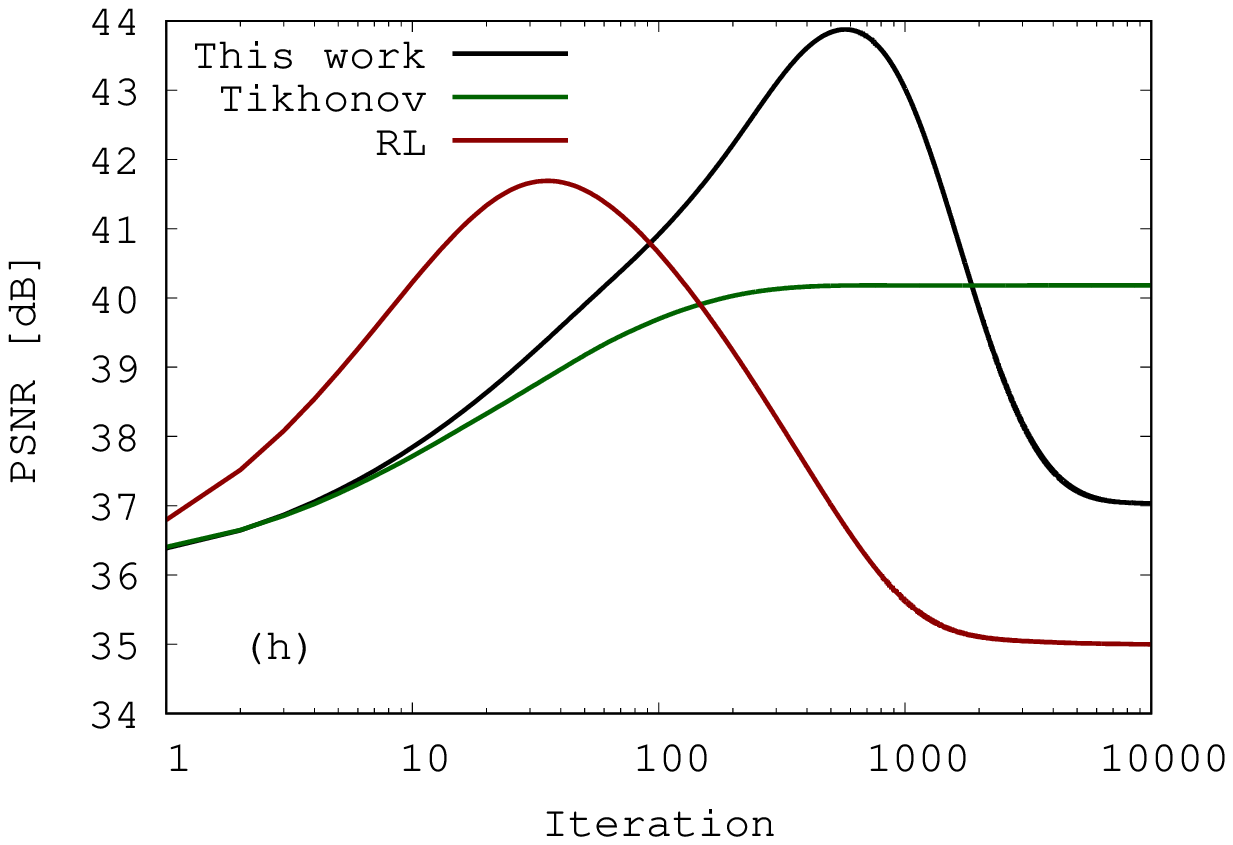}
    }
  \end{minipage}
  \\
  \vspace{0.2cm}
  \caption{
    Cost functions and their components against iterations (left) and PSNR against iterations for the three methods (right). 
    Panels from top to bottom indicate Spiral-1 with 5.0\% noise, Elliptical-1 with 1.0\% noise, Spiral-2 from the HSC-DUD, and Elliptical-2 from the HSC-Wide. 
    \label{fig:costs}}
\end{figure*}

The left panels of figure \ref{fig:costs} show objective functions with the components against the number of iterations for the four galaxies.
In all cases, the total cost and its components seem to converge at about 100 iterations.
On the other hand, the PSNR for our method continues to increase until the number of iterations reaches to 1000$\sim$10000, as shown in the black lines in the right panels of figure \ref{fig:costs}.
\par
The result that the PSNRs continues to increase despite the convergence of the objective function can be interpreted as the PSNR is sensitive to a small difference in the restored image. 
This is because the PSNR is expressed with a logarithmic of the MSE as shown in equation (\ref{eq:psnr}). 
In other words, the PSNR is sensitive to the detailed structures after the objective function almost converges and global structures are reproduced. 
It may lead to an idea of including a logarithmic function into the objective function to accelerate the convergence speed. 
However, it makes the calculation of the Lipschitz constant difficult and may also make the convergence unstable. 
Fortunately, our analysis shows that even the moderate number of iterations (i.e. $\sim$1000) produces a good image quality as shown in figure \ref{fig:costs} and tables \ref{tab:evasim}--\ref{tab:evaobs}. 
Hence, the conclusion that our regularization method is effective for deblurring galaxy images is robust. 

\par
\begin{figure*}
  \begin{minipage}{0.24 \hsize}
    \center{\includegraphics[width=80pt]{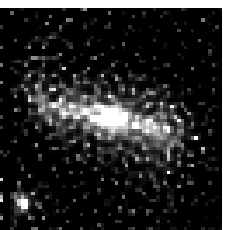}\\
      (a) RL\\
      (N=1000)}
  \end{minipage}
  \begin{minipage}{0.24 \hsize}
    \center{\includegraphics[width=80pt]{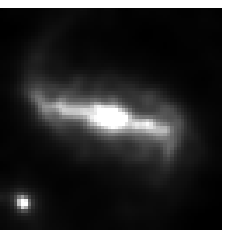}\\
      (b) This work\\
      (N=10000)}
  \end{minipage}
  \begin{minipage}{0.24 \hsize}
    \center{\includegraphics[width=80pt]{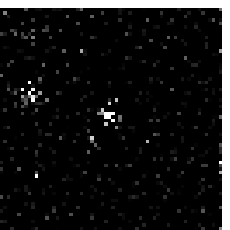}\\
      (c) RL\\
      (N=10000)}
  \end{minipage}
  \begin{minipage}{0.24 \hsize}
    \center{\includegraphics[width=80pt]{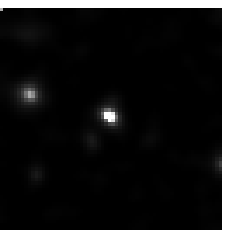}\\
      (d) This work\\
      (N=10000)}
  \end{minipage}
  \\
  \vspace{0.2cm}
  \caption{Comparison of the restored Spiral-1 with 1.0\% noise (a,b) and Elliptical-2 from the HSC-Wide (c,d) images with many iterations.
    \label{fig:highN}}
\end{figure*}

We apply the RL method with the optimal number of iterations that leads to the highest PSNR and SSIM.
As mentioned earlier, it is usually not possible to know the optimal number, and one has to stop the iteration at some time. 
However, the optimal number of iterations for the RL method significantly depends on cases. 
As shown in the red lines in the right panels of figure \ref{fig:costs}, it was 23, 34, 127, and 35 for Spiral-1 with 5.0\% noise, Elliptical-1 with 1.0\% noise, Spiral-2 from the HSC-DUD, and Elliptical-2 from the HSC-Wide, respectively. 
The serious problem is an over-optimization, which leads to a significantly low PSNR and degrades the image as shown in figure \ref{fig:highN}a and \ref{fig:highN}c. 
\par

\begin{figure*}
  \begin{minipage}{0.49 \hsize}
    \includegraphics[width=0.8\hsize]{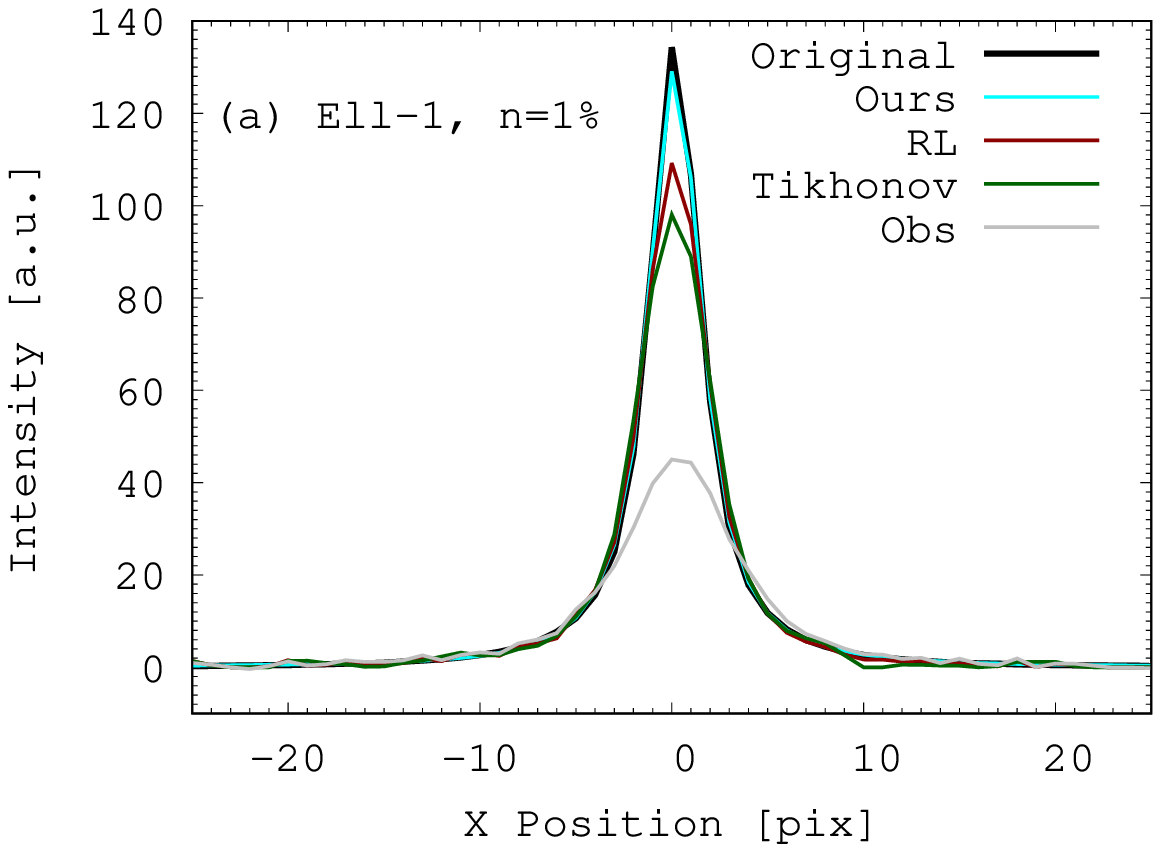}
  \end{minipage}
  \begin{minipage}{0.49 \hsize}
    \includegraphics[width=0.8\hsize]{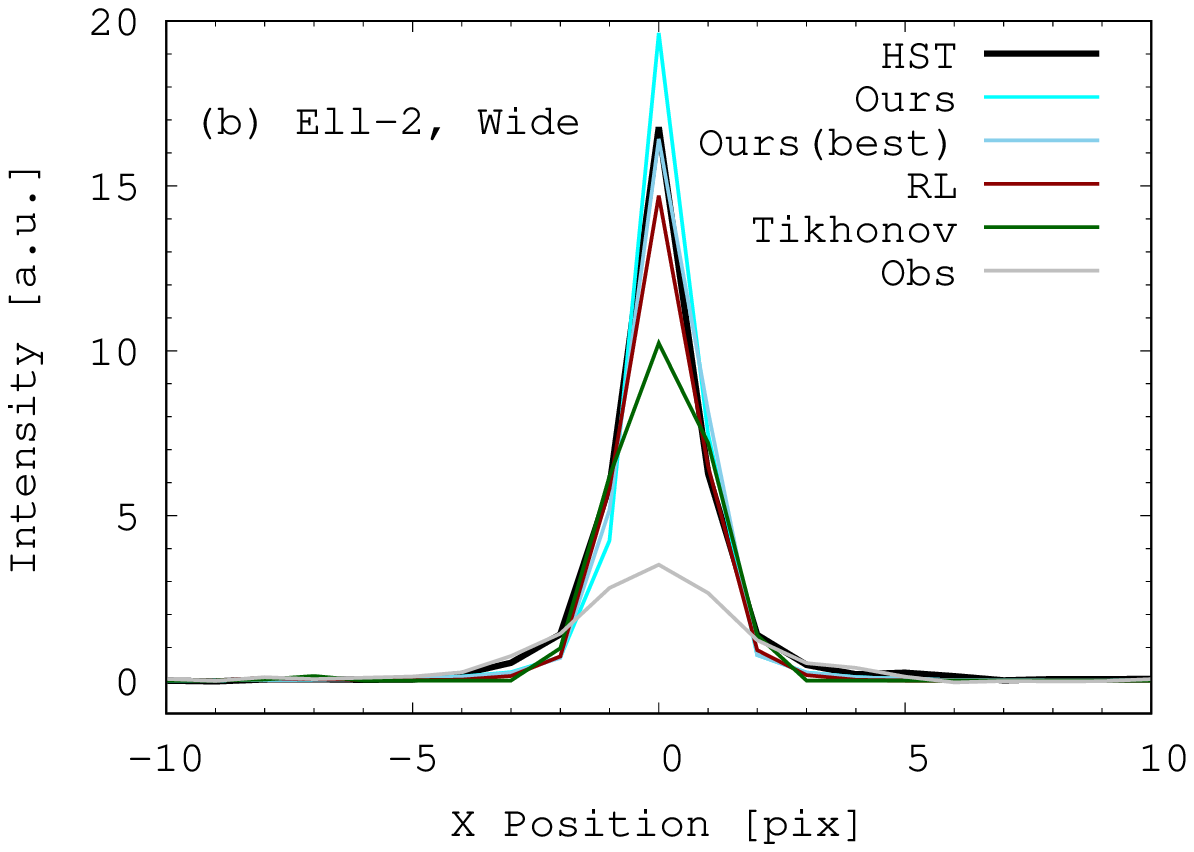}
  \end{minipage}
  \caption{Horizontal profiles of original, observed and restored images of Elliptical-1 with 5.0\% noise (a) and Elliptical-2 from the HSC-Wide (b).
    The faint blue line in the panel (b) indicates the result of our method with 572 iterations, which produces the best PSNR. 
    \label{fig:ellprf}}
\end{figure*}

In contrast, our method is stable during the iterations. 
As shown in figure \ref{fig:highN}b and \ref{fig:highN}d, the large number of iteration does not degrade the restored images. 
We note that the PSNR of our method decreases with the number of iterations for real observation images as shown in figure \ref{fig:costs}g and \ref{fig:costs}h.
It suggests that the best number of iterations is 572 for Elliptical-2 from HSC-Wide, and the adopted number of 1000 is too large. 
Although PSNRs from both RL and our methods decrease with increasing the iterations, the causes are different as indicated from figure \ref{fig:highN}. 
To explore the cause of the decrease in PSNR, we show the horizontal profile of Elliptical-2 from the HSC-Wide in figure \ref{fig:ellprf}b. 
It shows that, in our method, increasing the number of iterations leads to a sharper profile than the HST image. 
One possible interpretation is that the PSF is slightly fatter than the reality because this behaviour is only seen in case of real images. 
If this is the case, the optimization leads to too sharp images and it might be better to adopt a moderate number of iterations. 
Nonetheless, as indicated in figure \ref{fig:highN}b and  \ref{fig:highN}d, too many iterations do not lead to noise amplification. 
Hence, we conclude that our method can produce stable solutions.

\subsection{Dependence on hyper-parameters}
\label{hpar}
The qualities of restored images depend on hyper parameters and our method have four parameters: $b$, $\lambda$, $\tau$ and $\sigma$.
The latter two parameters, $\tau$ and $\sigma$, are the step lengths of the primal-dual splitting method and, in this study, they are automatically determined by equations (\ref{eq:tau}) and (\ref{eq:sig}).
In the following, we discuss the general characteristics of the former parameters, $b$ and $\lambda$, and investigate the dependence of our method on these parameters. 
\par

The parameter $b$ is used in the definition of asinh magnitude which is proportional to $x$ and $\log{x}$ when $|x| \gg b$ and $|x| \ll b$, respectively. 
The concept of using asinh magnitude is to appropriately treat pixels having zero or negative flux caused by a noise.
In this context, it is natural to choose $b$ to be related to flux uncertainty as suggested in \citet{1999AJ....118.1406L}.
In the regularization point of view, the parameter $b$ should be less than the brightness which fine structures to be restored would have. 
As discussed in section \ref{method}, galaxies have much steeper profiles than linear and this is why we apply asinh magnitude in our regularization. 
Nonetheless, galaxy structures fainter than $b$ is calculated in linear space and they can be smoothed out if $b$ is inappropriately large. 
In the optimization point of view, it is favourable that $b$ is proportional to the flux uncertainty. 
In this case, both the data-fidelity and regularization terms in our objective function (\ref{eq:cost}) include the square of flux uncertainty in the denominator and balancing these terms becomes less sensitive to S/N of the measurements. 
\par

The parameter $\lambda$ is used to balance the data fidelity and regularization terms in equation (\ref{eq:cost}).
In general, a higher $\lambda$ leads to smoothing out detailed structures while reducing noises, and this parameter should be chosen to balance these two effects.
As mentioned in the previous paragraph, both terms in our cost function include the square of flux uncertainty in the denominator. 
In this case, the balance between the data fidelity and regularization terms is less sensitive to $\lambda$. 
Exception is the case where some structures have a remarkably high S/N. 
The asinh magnitude in our regularization shows a logarithmic behaviour at higher S/N regions, leading to less weight to the regularization than the data fidelity. 
That is, high S/N images may favour high $\lambda$, contrary to intuition.
\par

\begin{figure*}[t]
  \begin{minipage}{0.99 \hsize}
    \includegraphics[width=0.49\hsize]{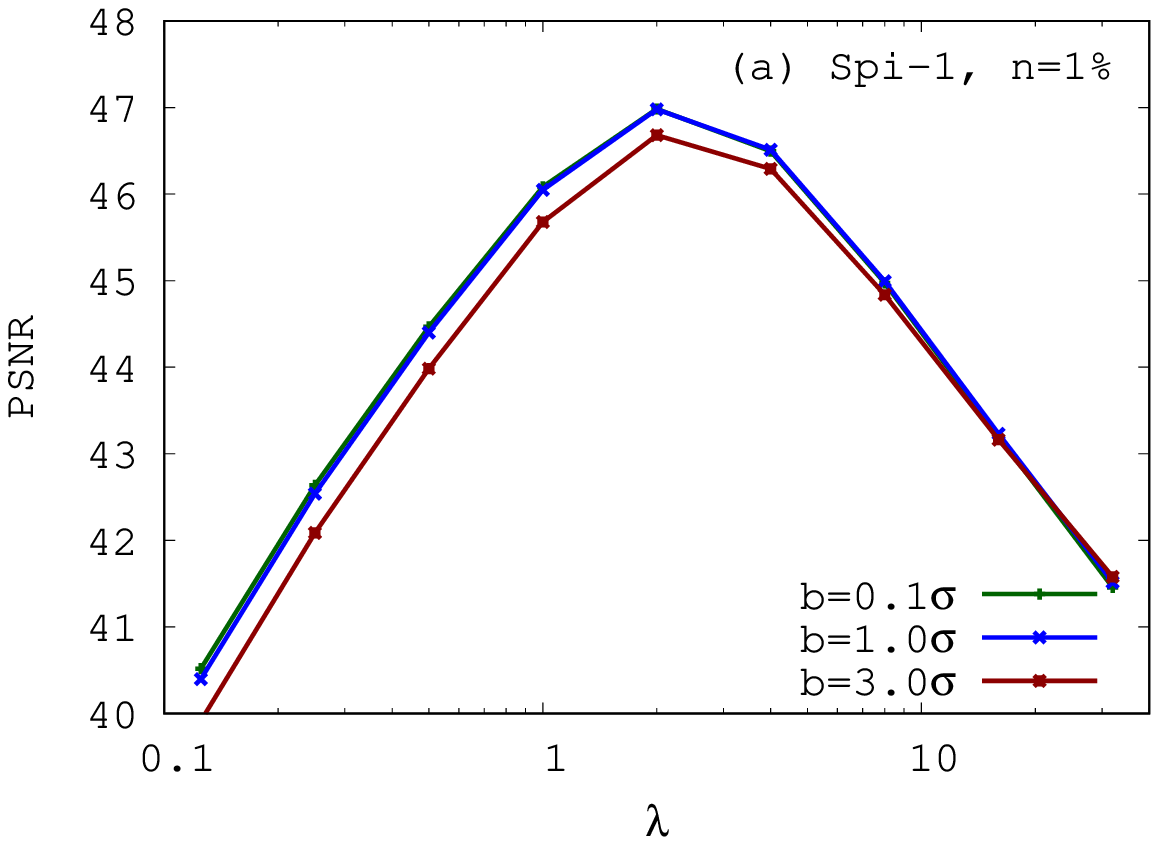}
    \includegraphics[width=0.49\hsize]{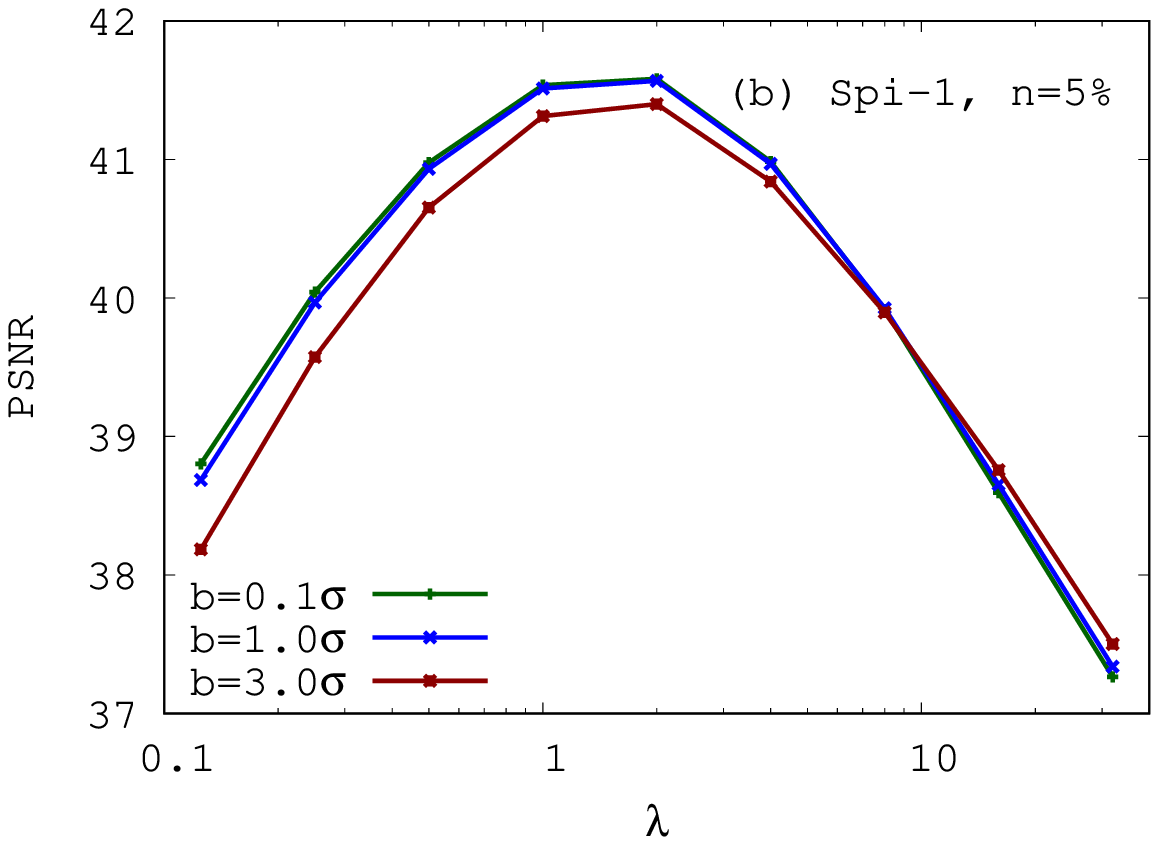}
  \end{minipage}
  \begin{minipage}{0.99 \hsize}
    \includegraphics[width=0.49\hsize]{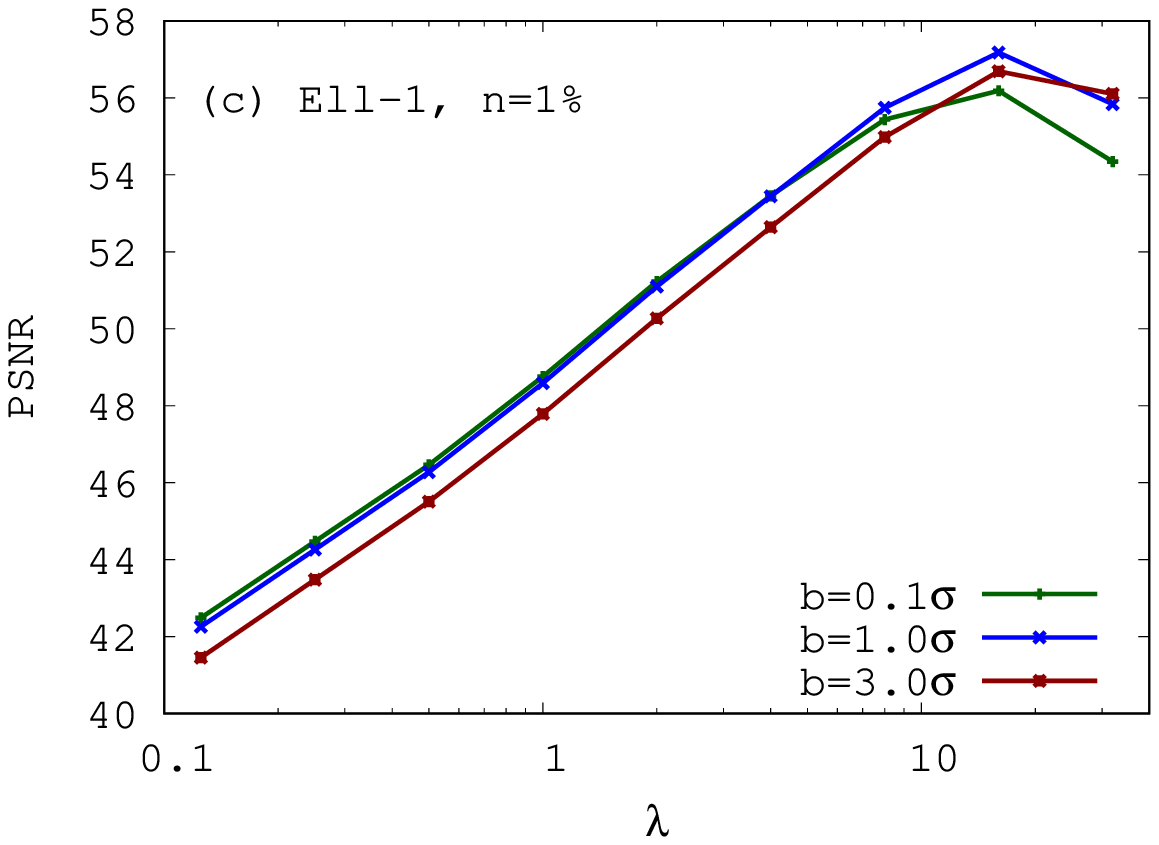}
    \includegraphics[width=0.49\hsize]{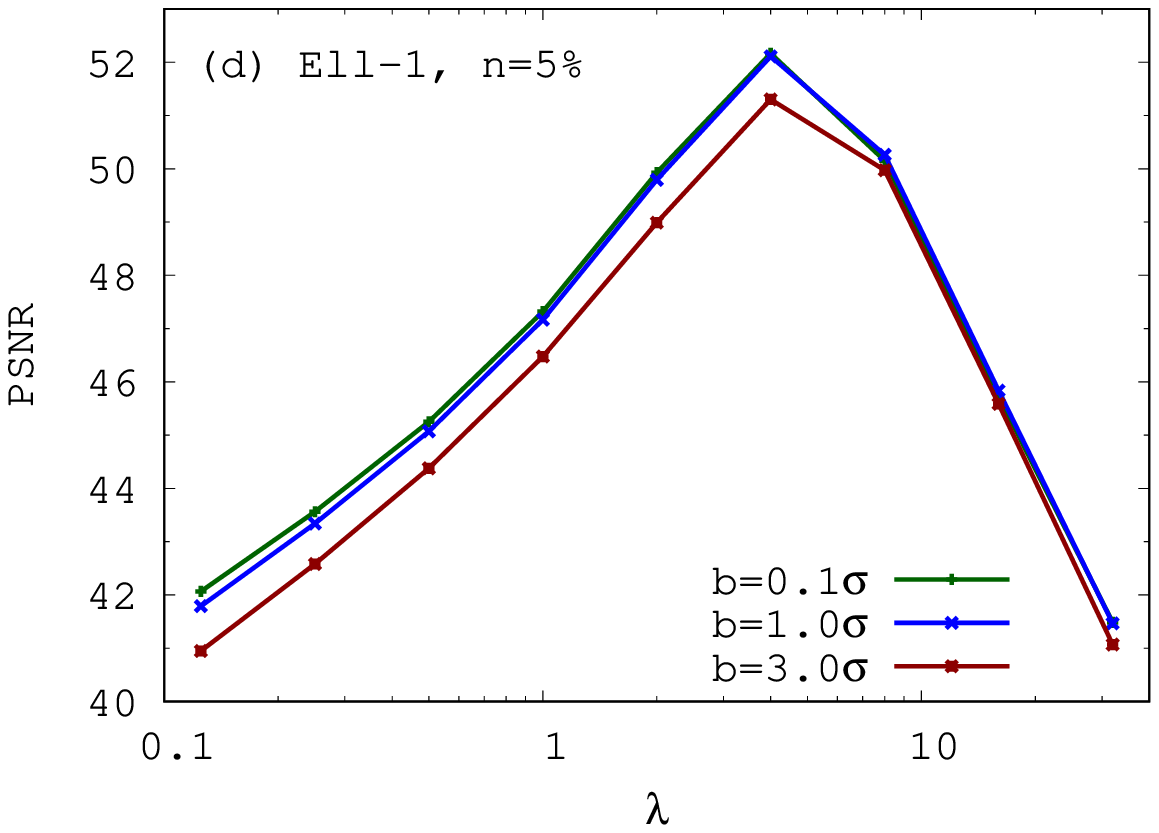}
  \end{minipage}
  \caption{
    The dependence on hyper-parameter, $\lambda$ and $b$, for PSNRs of restored images in the simulation. 
    The number of iterations is fixed to 10000 for a fare comparison. 
    \label{fig:hpar}}
\end{figure*}

We investigated the dependence of restored image quality on the two hyper parameters, $b$ and $\lambda$, via simulations.
We used Spiral-1 and Elliptical-1 with noises of 1\% and 5\%, the same as in section \ref{quality}.
The images were restored with our method with $\lambda=0.125-32$ and $b=0.05-3.0\, \bar{\sigma}_{y}$. 
The number of iterations was fixed to 10000. 
Figure \ref{fig:hpar} shows the PSNRs at various values of $\lambda$ and $b$ for the above galaxies. 
\par

It can be seen that the PSNR is not sensitive to {\it b} at $b<1.0\, \bar{\sigma}_{y}$ while a higher {\it b} leads to lower PSNRs. 
The latter indicates that fainter structures than {\it b} are smoothed out owing to unsuitable {\it b} as they can be treated in linear space as previously mentioned. 
From these results, we chose $b=1.0\, \bar{\sigma}_{y}$ throughout this paper. 
\par

Next, we discuss the dependence on $\lambda$, which is not as simple as $b$. 
The figure \ref{fig:hpar} shows that the best $\lambda$ is typically $2\sim 4$, i.e. the lower and higher $\lambda$ lead to noisy and blurred images, respectively.
On the other hand, for Elliptical-1 with 1\% noises, $\lambda=16$ leads to the highest PSNR. 
It can be interpreted as this galaxy has a steep profile and high S/N regions, which leads to less weight in the regularization term owing to the asinh magnitude. 
In this case, despite using flux uncertainty in the regularization term, the best regularization parameter still depends on the S/N.
\par
\begin{figure}
  \begin{minipage}{0.99 \hsize}
    \center{\includegraphics[width=90mm]{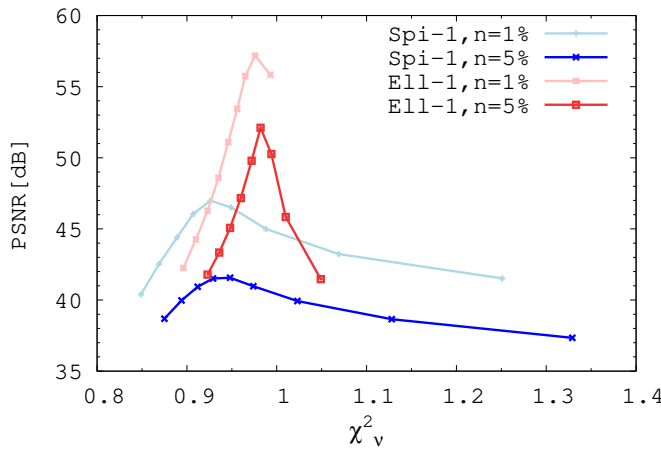}
    }
  \end{minipage}
  \caption{
    The PSNR for the objects in figure \ref{fig:hpar} against $\chi_\nu^2$.
    The larger $\chi_\nu^2$ corresponds to a larger $\lambda$.
    The parameter $b$ is fixed to 1.0 $\bar{\sigma}_{y}$.
    \label{fig:hparchi2}}
\end{figure}

 
On the other hand, the use of $\chi^{2}_{\nu}$ could mitigate the problem. 
Figure \ref{fig:hparchi2} shows the PSNR against $\chi^{2}_{\nu}$. 
A higher $\chi^{2}_{\nu}$ corresponds to lower weights in the data-fidelity term due to higher $\lambda$. 
We can see that the PSNR achieves the highest value at $\chi^{2}_{\nu}$ of slightly lower than 1.0, regardless of the galaxy type and the noise level.
Hence, $\lambda$ can be chosen to obtain $\chi^{2}_{\nu}$ of slightly less than 1.



\begin{table}
  \caption{
    Quantitative comparison of restored image qualities in the simulation. 
    \label{tab:evasim}}
  \begin{tabular}{ccccc}
    \hline
    \multirow{2}{*}{Spiral-1} & \multicolumn{2}{c}{$n=1\%$} & \multicolumn{2}{c}{$n=5\%$}\\
      & PSNR & SSIM & PSNR & SSIM \\
    \hline
    Blurred  & 35.47 & 0.850 & 35.28 & 0.845\\
    RL(max)   & 40.39 & 0.962 & 38.74 & 0.943\\
    Tikhonov  & 38.90 & 0.946 & 37.09 & 0.915\\
    This work      & {\bf 43.07} & {\bf 0.980} & {\bf 39.55} & {\bf 0.952}\\
    \hline
    \multirow{2}{*}{Elliptical-1} & \multicolumn{2}{c}{$n=1\%$} & \multicolumn{2}{c}{$n=5\%$}\\
    & PSNR & SSIM & PSNR & SSIM \\
    \hline
    Blurred  & 34.14 & 0.836 & 34.06 & 0.833 \\
    RL(max)   & 43.98 & 0.986 & 42.20 & 0.979 \\
    Tikhonov  & 41.69 & 0.979 & 38.15 & 0.948 \\
    This work      & {\bf 57.11} & {\bf 0.999} & {\bf 45.47} & {\bf 0.991} \\
    \hline
  \end{tabular}
\end{table}

\begin{table}
  \caption{The same as table \ref{tab:evasim} but for the observed images.
  }
  \label{tab:evaobs}
  \begin{tabular}{ccccc}
    \hline
    \multirow{2}{*}{Spiral-2} & \multicolumn{2}{c}{HSC-DUD} & \multicolumn{2}{c}{HSC-Wide}\\
      & PSNR & SSIM & PSNR & SSIM \\
    \hline
    Observed  & 34.00 & 0.939 & 35.51 & 0.959 \\
    RL(max)   & 40.98 & {\bf 0.990} & 39.69 & 0.986 \\
    Tikhonov  & 40.21 & 0.988 & 38.16 & 0.981 \\
    This work & {\bf 41.07} & {\bf 0.990} & {\bf 40.25} & {\bf 0.988} \\
    \hline
    \multirow{2}{*}{Elliptical-2} & \multicolumn{2}{c}{HSC-DUD} & \multicolumn{2}{c}{HSC-Wide}\\
    & PSNR & SSIM & PSNR & SSIM \\
    \hline
    Observed  & 34.75 & 0.420 & 35.64 & 0.554 \\
    RL(max)   & {\bf 44.25} & {\bf 0.962} & 41.69 & 0.872 \\
    Tikhonov  & 38.59 & 0.824 & 40.17 & 0.886 \\
    This work & 41.38 & 0.920 & {\bf 42.92} & {\bf 0.951} \\
    \hline
  \end{tabular}
\end{table}

\subsection{Limitation and future prospect}
In section \ref{quality}, we demonstrated that the proposed regularization, which minimizes image gradients in magnitude domain, leads to the best restored image quality.
Nonetheless, our method still has some limitations, which we discuss in the following. 
\par

The most considerable drawback of our method is the slow convergence; our method often requires a large number of iterations to obtain the best image quality, as discussed in section \ref{convergence}. 
Although images with a high quality can still be obtained with a moderate number of iterations, N=100-1000, more iterations are required to obtain the highest PSNR (figure \ref{fig:costs} e-h).
Calculation time under KM's computer setup (Intel Core i5, quad-core 1.1 GHz) was 15s and 180s for 1000 and 10000 iterations, respectively.
One thing to be considered is the rapid decrease of the cost function (figure \ref{fig:costs} a-d). 
It indicates that the global structures are rapidly reproduced while the local detailed structures are not because it is less sensitive to the objective function, as discussed in section \ref{convergence}. 
Modifying the cost function to be sensitive to the detailed structures may accelerate the convergence, which could be a future work.

\par
As in many other deblurring methods, our method provides only maximum-point estimation of the posterior probability distribution and no covariance matrix, although astronomical analysis usually requires uncertainty in pixel values. 
Besides, the variance could not be estimated from the restored image, since the pixel values are not independent of each other. 
Hence, some artefacts might be regarded as a ``significant detection''.
To address this problem, a Monte-Carlo technique or a bootstrap method with a different set of raw images could be a solution. 
A variational Bayes method could also be useful (e.g. \cite{Babacan08}, \cite{Ruiz15}, \cite{Sonogashira17}, \yearcite{Sonogashira18}). 
In this method, not only the best point in each pixel value, but also the probability distribution can be simultaneously estimated.
Although both the cost function and optimization method should be changed and the computational cost increases, this method is worth devising in a future work. 

\par
Finally, simultaneous restoration of multi-band images can be considered to produce resolution-matched images. 
To produce such data set, sharper images are usually blurred to match with the worst case, which leads to a loss of information and is not appropriate.  
However, deblurring all the images does not guarantee spatial resolution matching, especially in case of a different PSF size and/or S/N. 
To address this problem, the fact that galaxy structures are very similar in different band images can be used as {\it a priori} information. 
Of course the structures are not identical between different band images, but the colour gradients (in magnitude domain) should be smooth.
If this is the case, our method can naturally be extended towards the colour space, which could also be a future work. 

\section{Conclusion}
In this paper, we demonstrated the performance of Tikhonov regularization on magnitude domain for deblurring galaxy images.
The regularization on magnitude domain was a natural consequence of handling galaxy images, which have an exponential or S{\'e}rsic intensity profile. 
The objective function with images on both flux and magnitude domain was optimized by updating them separately.
We investigated the capability of our method via simulation and real observation images. 
The results showed that our method remarkably restored the spatial resolution of significantly blurred galaxy images. 
The quantitative evaluation confirmed that the proposed method was superior to both the classical RL method and conventional regularization-based method.

\begin{ack}
  This work has been supported by the Japan Society for the Promotion of Science (JSPS) Grants-in-Aid for Scientific Research (19H05076 and 21H01128). 
This work has also been supported in part by the Sumitomo Foundation Fiscal 2018 Grant for Basic Science Research Projects (180923), and the Collaboration Funding of the Institute of Statistical Mathematics ``New Development of the Studies on Galaxy Evolution with a Method of Data Science''. 
\par
We thank Prof. Shiro Ikeda for helpful discussions on this research.
\par
The Hyper Suprime-Cam (HSC) collaboration includes the astronomical communities of Japan and Taiwan, and Princeton University. The HSC instrumentation and software were developed by the National Astronomical Observatory of Japan (NAOJ), the Kavli Institute for the Physics and Mathematics of the Universe (Kavli IPMU), the University of Tokyo, the High Energy Accelerator Research Organization (KEK), the Academia Sinica Institute for Astronomy and Astrophysics in Taiwan (ASIAA), and Princeton University. Funding was contributed by the FIRST program from Japanese Cabinet Office, the Ministry of Education, Culture, Sports, Science and Technology (MEXT), the Japan Society for the Promotion of Science (JSPS), Japan Science and Technology Agency (JST), the Toray Science Foundation, NAOJ, Kavli IPMU, KEK, ASIAA, and Princeton University.
\par 
The Pan-STARRS1 Surveys (PS1) have been made possible through contributions of the Institute for Astronomy, the University of Hawaii, the Pan-STARRS Project Office, the Max-Planck Society and its participating institutes, the Max Planck Institute for Astronomy, Heidelberg and the Max Planck Institute for Extraterrestrial Physics, Garching, The Johns Hopkins University, Durham University, the University of Edinburgh, Queen's University Belfast, the Harvard-Smithsonian Center for Astrophysics, the Las Cumbres Observatory Global Telescope Network Incorporated, the National Central University of Taiwan, the Space Telescope Science Institute, the National Aeronautics and Space Administration under Grant No. NNX08AR22G issued through the Planetary Science Division of the NASA Science Mission Directorate, the National Science Foundation under Grant No. AST-1238877, the University of Maryland, and Eotvos Lorand University (ELTE) and the Los Alamos National Laboratory.
\par

\end{ack}

\bibliographystyle{aa}
\bibliography{deconv, imgana}

\begin{thebibliography}{38}
\expandafter\ifx\csname natexlab\endcsname\relax\def\natexlab#1{#1}\fi

\bibitem[{{Aihara} {et~al.}(2019){Aihara}, {AlSayyad}, {Ando}, {Armstrong},
  {Bosch}, {Egami}, {Furusawa}, {Furusawa}, {Goulding}, {Harikane}, {Hikage},
  {Ho}, {Hsieh}, {Huang}, {Ikeda}, {Imanishi}, {Ito}, {Iwata}, {Jaelani},
  {Kakuma}, {Kawana}, {Kikuta}, {Kobayashi}, {Koike}, {Komiyama}, {Li},
  {Liang}, {Lin}, {Luo}, {Lupton}, {Lust}, {MacArthur}, {Matsuoka}, {Mineo},
  {Miyatake}, {Miyazaki}, {More}, {Murata}, {Namiki}, {Nishizawa}, {Oguri},
  {Okabe}, {Okamoto}, {Okura}, {Ono}, {Onodera}, {Onoue}, {Osato}, {Ouchi},
  {Shibuya}, {Strauss}, {Sugiyama}, {Suto}, {Takada}, {Takagi}, {Takata},
  {Takita}, {Tanaka}, {Terai}, {Toba}, {Uchiyama}, {Utsumi}, {Wang}, {Wang}, \&
  {Yamada}}]{2019PASJ...71..114A}
{Aihara}, H., {et~al.} 2019, \pasj, 71, 114

\bibitem[{{Aihara} {et~al.}(2018){Aihara}, {Arimoto}, {Armstrong}, {Arnouts},
  {Bahcall}, {Bickerton}, {Bosch}, {Bundy}, {Capak}, {Chan}, {Chiba}, {Coupon},
  {Egami}, {Enoki}, {Finet}, {Fujimori}, {Fujimoto}, {Furusawa}, {Furusawa},
  {Goto}, {Goulding}, {Greco}, {Greene}, {Gunn}, {Hamana}, {Harikane},
  {Hashimoto}, {Hattori}, {Hayashi}, {Hayashi}, {He{\l}miniak}, {Higuchi},
  {Hikage}, {Ho}, {Hsieh}, {Huang}, {Huang}, {Ikeda}, {Imanishi}, {Inoue},
  {Iwasawa}, {Iwata}, {Jaelani}, {Jian}, {Kamata}, {Karoji}, {Kashikawa},
  {Katayama}, {Kawanomoto}, {Kayo}, {Koda}, {Koike}, {Kojima}, {Komiyama},
  {Konno}, {Koshida}, {Koyama}, {Kusakabe}, {Leauthaud}, {Lee}, {Lin}, {Lin},
  {Lupton}, {Mandelbaum}, {Matsuoka}, {Medezinski}, {Mineo}, {Miyama},
  {Miyatake}, {Miyazaki}, {Momose}, {More}, {More}, {Moritani}, {Moriya},
  {Morokuma}, {Mukae}, {Murata}, {Murayama}, {Nagao}, {Nakata}, {Niida},
  {Niikura}, {Nishizawa}, {Obuchi}, {Oguri}, {Oishi}, {Okabe}, {Okamoto},
  {Okura}, {Ono}, {Onodera}, {Onoue}, {Osato}, {Ouchi}, {Price}, {Pyo}, {Sako},
  {Sawicki}, {Shibuya}, {Shimasaku}, {Shimono}, {Shirasaki}, {Silverman},
  {Simet}, {Speagle}, {Spergel}, {Strauss}, {Sugahara}, {Sugiyama}, {Suto},
  {Suyu}, {Suzuki}, {Tait}, {Takada}, {Takata}, {Tamura}, {Tanaka}, {Tanaka},
  {Tanaka}, {Tanaka}, {Terai}, {Terashima}, {Toba}, {Tominaga}, {Toshikawa},
  {Turner}, {Uchida}, {Uchiyama}, {Umetsu}, {Uraguchi}, {Urata}, {Usuda},
  {Utsumi}, {Wang}, {Wang}, {Wong}, {Yabe}, {Yamada}, {Yamanoi}, {Yasuda},
  {Yeh}, {Yonehara}, \& {Yuma}}]{2018PASJ...70S...4A}
{Aihara}, H., {et~al.} 2018, \pasj, 70, S4

\bibitem[{{Akiyama} {et~al.}(2017){Akiyama}, {Ikeda}, {Pleau}, {Fish},
  {Tazaki}, {Kuramochi}, {Broderick}, {Dexter}, {Mo{\'s}cibrodzka},
  {Gowanlock}, {Honma}, \& {Doeleman}}]{2017AJ....153..159A}
{Akiyama}, K. {et~al.} 2017, \aj, 153, 159

\bibitem[{{Anconelli} {et~al.}(2007){Anconelli}, {Bertero}, {Boccacci},
  {Carbillet}, \& {Lanterib}}]{Anconelli04}
{Anconelli}, B., {Bertero}, M., {Boccacci}, P., {Carbillet}, M., \& {Lanterib},
  H. 2007, Journal of Computational and Applied Mathematics, 198, 321

\bibitem[{{Babacan} {et~al.}(2008){Babacan}, {Molina}, \&
  {Katsaggelos}}]{Babacan08}
{Babacan}, S.~D., {Molina}, R., \& {Katsaggelos}, A.~K. 2008, IEEE Transactions
  on Image Processing, 17, 326

\bibitem[{{Bauschke} {\& Combettes}(2011)}]{Bauschke11}
  Bauschke, H. H. \& Combettes, P. L. 2011, Convex Analysis and Monotone Operator Theory in Hilbert Spaces (New York: Springer), chap. 14 and 23
  
\bibitem[{{Boyd} {et~al.}(2011){Boyd}, N., {Chu}, B., \& J.}]{ADMM_Boyd2011}
{Boyd}, S., {Parikh.} N., {Chu}, E., {Peleato} B., \& {Eckstein} J., 2011, Foundations and Trends
  in Machine Learning, 3, 1

\bibitem[{{Chung} {et~al.}(2021){Chung}, {Park}, \&
  {Park}}]{2021ApJS..257...66C}
{Chung}, H., {Park}, C., \& {Park}, Y.-S. 2021, \apjs, 257, 66

\bibitem[{{Condat}(2013)}]{Condat2013}
{Condat}, L. 2013, J. Optimization Theory and Applications, 158, 460

\bibitem[{{Event Horizon Telescope Collaboration} {et~al.}(2019){Event Horizon
  Telescope Collaboration}, {Akiyama}, {Alberdi}, {Alef}, {Asada}, {Azulay},
  {Baczko}, {Ball}, {Balokovi{\'c}}, {Barrett}, {Bintley}, {Blackburn},
  {Boland}, {Bouman}, {Bower}, {Bremer}, {Brinkerink}, {Brissenden}, {Britzen},
  {Broderick}, {Broguiere}, {Bronzwaer}, {Byun}, {Carlstrom}, {Chael}, {Chan},
  {Chatterjee}, {Chatterjee}, {Chen}, {Chen}, {Cho}, {Christian}, {Conway},
  {Cordes}, {Crew}, {Cui}, {Davelaar}, {De Laurentis}, {Deane}, {Dempsey},
  {Desvignes}, {Dexter}, {Doeleman}, {Eatough}, {Falcke}, {Fish}, {Fomalont},
  {Fraga-Encinas}, {Freeman}, {Friberg}, {Fromm}, {G{\'o}mez}, {Galison},
  {Gammie}, {Garc{\'\i}a}, {Gentaz}, {Georgiev}, {Goddi}, {Gold}, {Gu},
  {Gurwell}, {Hada}, {Hecht}, {Hesper}, {Ho}, {Ho}, {Honma}, {Huang}, {Huang},
  {Hughes}, {Ikeda}, {Inoue}, {Issaoun}, {James}, {Jannuzi}, {Janssen},
  {Jeter}, {Jiang}, {Johnson}, {Jorstad}, {Jung}, {Karami}, {Karuppusamy},
  {Kawashima}, {Keating}, {Kettenis}, {Kim}, {Kim}, {Kim}, {Kino}, {Koay},
  {Koch}, {Koyama}, {Kramer}, {Kramer}, {Krichbaum}, {Kuo}, {Lauer}, {Lee},
  {Li}, {Li}, {Lindqvist}, {Liu}, {Liuzzo}, {Lo}, {Lobanov}, {Loinard},
  {Lonsdale}, {Lu}, {MacDonald}, {Mao}, {Markoff}, {Marrone}, {Marscher},
  {Mart{\'\i}-Vidal}, {Matsushita}, {Matthews}, {Medeiros}, {Menten}, {Mizuno},
  {Mizuno}, {Moran}, {Moriyama}, {Moscibrodzka}, {M{\"u}ller}, {Nagai},
  {Nagar}, {Nakamura}, {Narayan}, {Narayanan}, {Natarajan}, {Neri}, {Ni},
  {Noutsos}, {Okino}, {Olivares}, {Ortiz-Le{\'o}n}, {Oyama}, {{\"O}zel},
  {Palumbo}, {Patel}, {Pen}, {Pesce}, {Pi{\'e}tu}, {Plambeck}, {PopStefanija},
  {Porth}, {Prather}, {Preciado-L{\'o}pez}, {Psaltis}, {Pu}, {Ramakrishnan},
  {Rao}, {Rawlings}, {Raymond}, {Rezzolla}, {Ripperda}, {Roelofs}, {Rogers},
  {Ros}, {Rose}, {Roshanineshat}, {Rottmann}, {Roy}, {Ruszczyk}, {Ryan},
  {Rygl}, {S{\'a}nchez}, {S{\'a}nchez-Arguelles}, {Sasada}, {Savolainen},
  {Schloerb}, {Schuster}, {Shao}, {Shen}, {Small}, {Sohn}, {SooHoo}, {Tazaki},
  {Tiede}, {Tilanus}, {Titus}, {Toma}, {Torne}, {Trent}, {Trippe}, {Tsuda},
  {van Bemmel}, {van Langevelde}, {van Rossum}, {Wagner}, {Wardle},
  {Weintroub}, {Wex}, {Wharton}, {Wielgus}, {Wong}, {Wu}, {Young}, {Young},
  {Younsi}, {Yuan}, {Yuan}, {Zensus}, {Zhao}, {Zhao}, {Zhu}, {Algaba},
  {Allardi}, {Amestica}, {Anczarski}, {Bach}, {Baganoff}, {Beaudoin}, {Benson},
  {Berthold}, {Blanchard}, {Blundell}, {Bustamente}, {Cappallo},
  {Castillo-Dom{\'\i}nguez}, {Chang}, {Chang}, {Chang}, {Chen}, {Chilson},
  {Chuter}, {C{\'o}rdova Rosado}, {Coulson}, {Crawford}, {Crowley}, {David},
  {Derome}, {Dexter}, {Dornbusch}, {Dudevoir}, {Dzib}, {Eckart}, {Eckert},
  {Erickson}, {Everett}, {Faber}, {Farah}, {Fath}, {Folkers}, {Forbes},
  {Freund}, {G{\'o}mez-Ruiz}, {Gale}, {Gao}, {Geertsema}, {Graham}, {Greer},
  {Grosslein}, {Gueth}, {Haggard}, {Halverson}, {Han}, {Han}, {Hao},
  {Hasegawa}, {Henning}, {Hern{\'a}ndez-G{\'o}mez}, {Herrero-Illana},
  {Heyminck}, {Hirota}, {Hoge}, {Huang}, {Impellizzeri}, {Jiang}, {Kamble},
  {Keisler}, {Kimura}, {Kono}, {Kubo}, {Kuroda}, {Lacasse}, {Laing}, {Leitch},
  {Li}, {Lin}, {Liu}, {Liu}, {Lu}, {Marson}, {Martin-Cocher}, {Massingill},
  {Matulonis}, {McColl}, {McWhirter}, {Messias}, {Meyer-Zhao}, {Michalik},
  {Monta{\~n}a}, {Montgomerie}, {Mora-Klein}, {Muders}, {Nadolski}, {Navarro},
  {Neilsen}, {Nguyen}, {Nishioka}, {Norton}, {Nowak}, {Nystrom}, {Ogawa},
  {Oshiro}, {Oyama}, {Parsons}, {Paine}, {Pe{\~n}alver}, {Phillips}, {Poirier},
  {Pradel}, {Primiani}, {Raffin}, {Rahlin}, {Reiland}, {Risacher}, {Ruiz},
  {S{\'a}ez-Mada{\'\i}n}, {Sassella}, {Schellart}, {Shaw}, {Silva}, {Shiokawa},
  {Smith}, {Snow}, {Souccar}, {Sousa}, {Sridharan}, {Srinivasan}, {Stahm},
  {Stark}, {Story}, {Timmer}, {Vertatschitsch}, {Walther}, {Wei}, {Whitehorn},
  {Whitney}, {Woody}, {Wouterloot}, {Wright}, {Yamaguchi}, {Yu}, {Zeballos},
  {Zhang}, \& {Ziurys}}]{2019ApJ...875L...1E}
{Event Horizon Telescope Collaboration}, {et~al.}
  2019, \apjl, 875, L1

\bibitem[{{F{\'e}tick} {et~al.}(2020){F{\'e}tick}, {Mugnier}, {Fusco}, \&
  {Neichel}}]{2020MNRAS.496.4209F}
{F{\'e}tick}, R.~J.~L., {Mugnier}, L.~M., {Fusco}, T., \& {Neichel}, B. 2020,
  \mnras, 496, 4209

\bibitem[{{Gan} {et~al.}(2021){Gan}, {Bekki}, \&
  {Hashemizadeh}}]{2021arXiv210309711G}
{Gan}, F.~K., {Bekki}, K., \& {Hashemizadeh}, A. 2021, arXiv e-prints,
  arXiv:2103.09711

\bibitem[{{Honma} {et~al.}(2014){Honma}, {Akiyama}, {Uemura}, \&
  {Ikeda}}]{2014PASJ...66...95H}
{Honma}, M., {Akiyama}, K., {Uemura}, M., \& {Ikeda}, S. 2014, \pasj, 66, 95

\bibitem[{{Hope} {et~al.}(2022){Hope}, {Jefferies}, {Li Causi}, {Landoni},
  {Stangalini}, {Pedichini}, \& {Antoniucci}}]{2022ApJ...926...88H}
  {Hope}, D.~A., {Jefferies}, S.~M., {Li Causi}, G., {Landoni}, M., {Stangalini}, M. and {Pedichini}, F. \& {Antoniucci}, S. 2022, \apj, 926, 88

\bibitem[{{Idier}(2008)}]{Idier}
{Idier}, J. 2008, Bayesian Approach to Inverse Problems (Wiley)

\bibitem[{{Jenkner} {et~al.}(2006){Jenkner}, {Doxsey}, {Hanisch}, {Lubow},
  {Miller}, \& {White}}]{2006ASPC..351..406J}
  {Jenkner}, H., {Doxsey}, R.~E., {Hanisch}, R.~J.,
  {Lubow}, S.~H., {Miller}, W.~W., III, \& {White}, R.~L., 2006, in
  Astronomical Society of the Pacific Conference Series, Vol. 351, Astronomical
  Data Analysis Software and Systems XV, ed. C.~{Gabriel}, C.~{Arviset},
  D.~{Ponz}, \& S.~{Enrique}, 406

\bibitem[{{Kuramochi} {et~al.}(2018){Kuramochi}, {Akiyama}, {Ikeda}, {Tazaki},
  {Fish}, {Pu}, {Asada}, \& {Honma}}]{2018ApJ...858...56K}
  {Kuramochi}, K., {Akiyama}, K., {Ikeda}, S.,
  {Tazaki}, F., {Fish}, V.~L., {Pu}, H., {Asada}, K. \& {Honma}, M.  2018, \apj, 858, 56

\bibitem[{{Lucy}(1974)}]{1974AJ.....79..745L}
{Lucy}, L.~B. 1974, \aj, 79, 745

\bibitem[{{Lupton} {et~al.}(1999){Lupton}, {Gunn}, \&
  {Szalay}}]{1999AJ....118.1406L}
{Lupton}, R.~H., {Gunn}, J.~E., \& {Szalay}, A.~S. 1999, \aj, 118, 1406

\bibitem[{{Nammour} {et~al.}(2022){Nammour}, {Akhaury}, {Girard}, {Lanusse},
  {Sureau}, {Ben Ali}, \& {Starck}}]{2022arXiv220307412N}
  {Nammour}, F., {Akhaury}, U., {Girard}, J.~N.,
  {Lanusse}, F. and {Sureau}, F. and {Ben Ali}, C. \& {Starck}, J. -L.  2022, arXiv e-prints,
  arXiv:2203.07412

\bibitem[{{Narayan} \& {Nityananda}(1986)}]{1986ARA&A..24..127N}
{Narayan}, R. \& {Nityananda}, R. 1986, \araa, 24, 127

\bibitem[{{Richardson}(1972)}]{1972JOSA...62...55R}
{Richardson}, W.~H. 1972, Journal of the Optical Society of America
  (1917-1983), 62, 55

\bibitem[{{Rudin} {et~al.}(1992){Rudin}, {Osher}, \&
  {Fatemi}}]{1992PhyD...60..259R}
{Rudin}, L.~I., {Osher}, S., \& {Fatemi}, E. 1992, Physica D Nonlinear
  Phenomena, 60, 259

\bibitem[{{Ruiz} {et~al.}(2015){Ruiz}, {Zhou}, {Mateos}, {Molina}, \&
  {Katsaggelos}}]{Ruiz15}
{Ruiz}, P., {Zhou}, X., {Mateos}, J., {Molina}, R., \& {Katsaggelos}, A.~K.
  2015, Digital Signal Processing, 47, 116

\bibitem[{{Rydbeck}(2008)}]{2008ApJ...675.1304R}
{Rydbeck}, G. 2008, \apj, 675, 1304

\bibitem[{{Schawinski} {et~al.}(2017){Schawinski}, {Zhang}, {Zhang}, {Fowler},
  \& {Santhanam}}]{2017MNRAS.467L.110S}
{Schawinski}, K., {Zhang}, C., {Zhang}, H., {Fowler}, L., \& {Santhanam}, G.~K.
  2017, \mnras, 467, L110

\bibitem[{{S{\'e}rsic}(1963)}]{1963BAAA....6...41S}
{S{\'e}rsic}, J.~L. 1963, Boletin de la Asociacion Argentina de Astronomia La
  Plata Argentina, 6, 41

\bibitem[{{Shi} {et~al.}(2017){Shi}, {Guo}, {Zhu}, \& {Wang}}]{Shi17}
{Shi}, X., {Guo}, R., {Zhu}, Y., \& {Wang}, Z. 2017, Journal of Systems
  Engineering and Electronics, 28, 1236

\bibitem[{{Shibuya} {et~al.}(2022){Shibuya}, {Miura}, {Iwadate}, {Fujimoto},
  {Harikane}, {Toba}, {Umayahara}, \& {Ito}}]{2022PASJ...74...73S}
  {Shibuya}, T., {Miura}, N., {Iwadate}, K.,
  {Fujimoto}, S., {Harikane}, Y., {Toba}, Y., {Umayahara}, T. \& {Ito}, Y. 
  2022, \pasj, 74, 73

\bibitem[{{Sonogashira} {et~al.}(2017){Sonogashira}, {Funatomi}, {Iiyama}, \&
  Michihiko}]{Sonogashira17}
{Sonogashira}, M., {Funatomi}, T., {Iiyama}, M., \& Michihiko, M. 2017, IEEE
  Transactions on Image Processing, 26

\bibitem[{{Sonogashira} {et~al.}(2018){Sonogashira}, {Funatomi}, {Iiyama}, \&
  Michihiko}]{Sonogashira18}
{Sonogashira}, M., {Funatomi}, T., {Iiyama}, M., \& Michihiko, M. 2018, IPSJ
  SIG Technical Report, 2018-CVIM-212

\bibitem[{{Starck} {et~al.}(2007){Starck}, {Fadili}, \&
  {Murtagh}}]{2007ITIP...16..297S}
{Starck}, J.-L., {Fadili}, J., \& {Murtagh}, F. 2007, IEEE Transactions on
  Image Processing, 16, 297

\bibitem[{{Starck} {et~al.}(2002){Starck}, {Pantin}, \&
  {Murtagh}}]{2002PASP..114.1051S}
{Starck}, J.~L., {Pantin}, E., \& {Murtagh}, F. 2002, \pasp, 114, 1051

\bibitem[{{Sureau} {et~al.}(2020){Sureau}, {Lechat}, \&
  {Starck}}]{2020A&A...641A..67S}
{Sureau}, F., {Lechat}, A., \& {Starck}, J.~L. 2020, \aap, 641, A67

\bibitem[{{Suzuki} {et~al.}(2019){Suzuki}, {Minowa}, {Koyama}, {Kodama},
  {Hayashi}, {Shimakawa}, {Tanaka}, \& {Tadaki}}]{2019PASJ...71...69S}
  {Suzuki}, T.~L., {Minowa}, Y., {Koyama}, Y.,
  {Kodama}, T. {Hayashi}, M.. {Shimakawa}, R., {Tanaka}, I. \& {Tadaki}, K.
  2019, \pasj, 71, 69

\bibitem[{{Tanaka} {et~al.}(2018){Tanaka}, {Coupon}, {Hsieh}, {Mineo},
  {Nishizawa}, {Speagle}, {Furusawa}, {Miyazaki}, \&
  {Murayama}}]{2018PASJ...70S...9T}
  {Tanaka}, M., {et~al.} 2018, \pasj, 70, S9

\bibitem[{{Wang} {et~al.}(2004){Wang}, {Bovik}, {Sheikh}, \&
  {Simoncelli}}]{2004ITIP...13..600W}
{Wang}, Z., {Bovik}, A.~C., {Sheikh}, H.~R., \& {Simoncelli}, E.~P. 2004, IEEE
  Transactions on Image Processing, 13, 600

\bibitem[{{Whitmore} {et~al.}(2008){Whitmore}, {Lindsay}, \&
  {Stankiewicz}}]{2008ASPC..394..481W}
{Whitmore}, B., {Lindsay}, K., \& {Stankiewicz}, M. 2008, in Astronomical
  Society of the Pacific Conference Series, Vol. 394, Astronomical Data
  Analysis Software and Systems XVII, ed. R.~W. {Argyle}, P.~S. {Bunclark}, \&
  J.~R. {Lewis}, 481

\bibitem[{{Whitmore} {et~al.}(2016){Whitmore}, {Allam}, {Budav{\'a}ri},
  {Casertano}, {Downes}, {Donaldson}, {Fall}, {Lubow}, {Quick}, {Strolger},
  {Wallace}, \& {White}}]{2016AJ....151..134W}
{Whitmore}, B.~C., {et~al.} 2016, \aj, 151,
  134

\end{thebibliography}

\end{document}